\documentclass[11pt]{article}

\usepackage{amsmath,amssymb}

\usepackage{float}
\usepackage{pgf, tikz}
\usetikzlibrary{arrows, automata}

\usepackage{hyperref}
\usepackage[textwidth=8em,textsize=small]{todonotes}
\usepackage[utf8]{inputenc}
\usepackage{mathbbol}
\usepackage{bm}
\usepackage{url}
\usepackage{amsmath}
\usepackage{natbib}
\usepackage[utf8]{inputenc}
\usepackage{xcolor}
\usepackage{graphicx}
\usepackage{enumerate}
\usepackage{amsthm}
\usepackage{amsmath}
\usepackage{amssymb}
\usepackage{marginnote}
\usepackage{mathtools}
\usepackage{float}

\usepackage{longtable}
\usepackage{booktabs,array}
\usepackage{calc} 
\usepackage{etoolbox}
\usepackage{colortbl}

\usepackage{changepage}
\usepackage{color}
\usepackage{graphicx}
\usepackage{setspace}
\usepackage{lscape}
\usepackage{algorithm}
\usepackage{algpseudocode}

\usepackage{etoolbox}
\makeatletter
\patchcmd{\@makecaption}
  {\parbox}
  {\advance\@tempdima-\fontdimen2} 
  {}{}
\makeatother  

\providecommand{\tightlist}{%
  \setlength{\itemsep}{0pt}\setlength{\parskip}{0pt}}

\usepackage[top=0.7in, bottom=0.7in, left=1in, right=1in]{geometry}

\begin{document}
\vspace*{0.2in}

\begin{center}
{\Large \textbf{Accounting for reporting delays in real-time phylodynamic analyses with preferential sampling}} \
\\ \ \\
Catalina M. Medina\textsuperscript{1}, Julia A.
Palacios\textsuperscript{2}, Volodymyr M. Minin\textsuperscript{1*}
\\
\bigskip
$^1$Department of Statistics, University of California,
Irvine, \\ $^2$Departments of Statistics and Biomedical Data
Science, Stanford University, 
\bigskip

$^*$vminin@uci.edu
\end{center}

\begin{abstract}
The COVID-19 pandemic demonstrated that fast and accurate analysis of
continually collected infectious disease surveillance data is crucial
for situational awareness and policy making. Coalescent-based
phylodynamic analysis can use genetic sequences of a pathogen to
estimate changes in its effective population size, a measure of genetic
diversity. These changes in effective population size can be connected
to the changes in the number of infections in the population of interest
under certain conditions. Phylodynamics is an important set of tools
because its methods are often resilient to the ascertainment biases
present in traditional surveillance data (e.g., preferentially testing
symptomatic individuals). Unfortunately, it takes weeks or months to
sequence and deposit the sampled pathogen genetic sequences into a
database, making them available for such analyses. These reporting
delays severely decrease precision of phylodynamic methods closer to
present time, and for some models can lead to extreme biases. Here we
present a method that affords reliable estimation of the effective
population size trajectory closer to the time of data collection,
allowing for policy decisions to be based on more recent data. Our work
uses readily available historic times between sampling and sequencing
for a population of interest, and incorporates this information into the
sampling model to mitigate the effects of reporting delay in real-time
analyses. We illustrate our methodology on simulated data and on
SARS-CoV-2 sequences collected in the state of Washington in 2021.
\end{abstract}

\section{Introduction}\label{introduction}

The COVID-19 pandemic demonstrated that fast and accurate analysis of
continually collected infectious disease surveillance data is crucial
for situational awareness and policy making
\citep{cori2024, engebretsen2023}. Phylodynamic methods form an
important set of tools that use genetic sequences of a pathogen of
interest to infer its phylogeny and parameters of disease dynamics, such
as the effective population size. The effective population size is a
measure of genetic diversity, and estimation of effective population
size is often of interest because under certain conditions this quantity
can be connected to the number of infections in the population
\citep{volz2009} or in some cases more directly to transmission
\citep{frost2010}. Inference of the effective population size can also
be useful to compare the growth of different viral lineages
\citep{fountain-jones2020, volz2021}, as one part of an argument for the
effectiveness of an intervention \citep{vanballegooijen2009}, and
ultimately, for informed health policy decisions \citep{rich2023}.

The COVID-19 pandemic resulted in a massive push towards sharing sampled
pathogenic sequences in public databases such as: GISAID
(\href{http://www.gisaid.org/}{www.gisaid.org}) , NCBI
(\href{http://www.ncbi.nlm.nih.gov/}{www.ncbi.nlm.nih.gov}), and ViPR
(\href{http://www.viprbrc.org/}{www.viprbrc.org}). Unfortunately,
collected samples can take weeks or even months to sequence and upload
to a database, making them available for analysis \citep{kalia2021}. We
refer to this time between sample collection and sequence reporting as
the \emph{reporting delay} for a sample. Reporting delays result in
missing data near present time since recently collected samples are less
likely to have been sequenced and uploaded yet. During the COVID-19
pandemic, reporting delays were a novel and important consideration to
most, with the emerging need for real-time analysis, i.e., analysis
conducted up to present time \citep{kalia2021}. The distribution of
delays can be location, time, and even lineage specific
\citep{petrone2023}, influenced by factors such as sequencing cost and
laboratory limited capacity. Researchers who had considered reporting
delays for surveillance data in real-time analyses, were limited to
methods that utilized only aggregated level reporting delay information
\citep{bastos2019}. The shared public databases of pathogenic sequences
provide a new opportunity to utilize detailed sequence-level data of
reporting delays.

Modern methods to estimate effective population size changes from
genetic data have evolved from the original coalescent skyline plot
where the effective population size trajectory, \(N_e(t)\), was modeled
nonparametrically as piecewise constant \citep{pybus2000}, to grouping
methods that resulted in smoother estimates \citep{strimmer2001}, to the
first Bayesian coalescent skyline plot model \citep{drummond2005} which
jointly inferred a pathogen's evolutionary tree and \(N_e(t)\). Several
advancements on the Bayesian coalescent skyline plot models have been
proposed in recent years which consider different interval
specifications for the piecewise \(N_e(t)\) or regularization methods
for \(N_e(t)\). See Billenstein and Höhna (2024) for a detailed
comparison of Bayesian nonparametric inference of \(N_e(t)\) methods
\citep{billenstein2024}. When pathogen samples are being continually
collected over time it is often the case that the frequency at which
samples are collected is related to the burden of the infection in the
population. This is known as preferential sampling, and Karcher et
al.~(2016) proposed a phylodynamic model that built on Bayesian
coalescent skyline plot models to relate the sampling intensity to the
effective population size \citep{karcher2016}. It was shown that
unaccounted for preferential sampling can result in biases and
accounting for preferential sampling can result in more accurate and
precise inference of the effective population size trajectory. This
model has been extended to allow for additional factors to be related to
the sampling intensity and effective population size
\citep{karcher2020, cappello2022}.

In this work we use simulations to investigate the effects of reporting
delays in real-time phylodynamic inference of the effective population
size; we compare the effects across various state-of-the-art inferential
strategies. We also propose a strategy to mitigate the effects of
reporting delays within the preferential model, by incorporating
information about the distribution of recent reporting delays. This
extends the Karcher et al.~(2020) model by including reporting
probabilities into the sampling intensity model \citep{karcher2020}. We
use simulations to compare the performance of our proposed model with
competitive real-time phylodynamic strategies in the presence of
preferential sampling and reporting delays and show that our model has
lower bias, better coverage, and higher precision than state-of-the-art
methods. Finally, we use SARS-CoV-2 sequences from Washington state as a
case study to compare real-time inferential strategies on data which
suffers from reporting delays to the performance of retrospective
inference on all sampled sequences in the hypothetical case of no
reporting delays.

\section{Methods}\label{methods}

We will begin with a description of the nonparametric phylodynamic
methodology proposed in Karcher et al.~(2020) \citep{karcher2020}. This
Bayesian strategy will be described starting with how the pathogen
genetic samples are modeled conditionally on its evolutionary tree,
sampling times, number of samples at each time, and effective population
size trajectory, followed by details of the overall full hierarchical
model. Once this framework is understood, we will introduce our proposal
to mitigate the effects of delays between collecting a sample and
depositing a pathogen sequence obtained from the sample into a public
database.

\subsection{\texorpdfstring{Summary of Bayesian nonparametric \(N_e(t)\)
inference}{Summary of Bayesian nonparametric N\_e(t) inference}}\label{summary-of-bayesian-nonparametric-n_et-inference}

\begin{figure}

\centering{

\includegraphics[width=\textwidth]{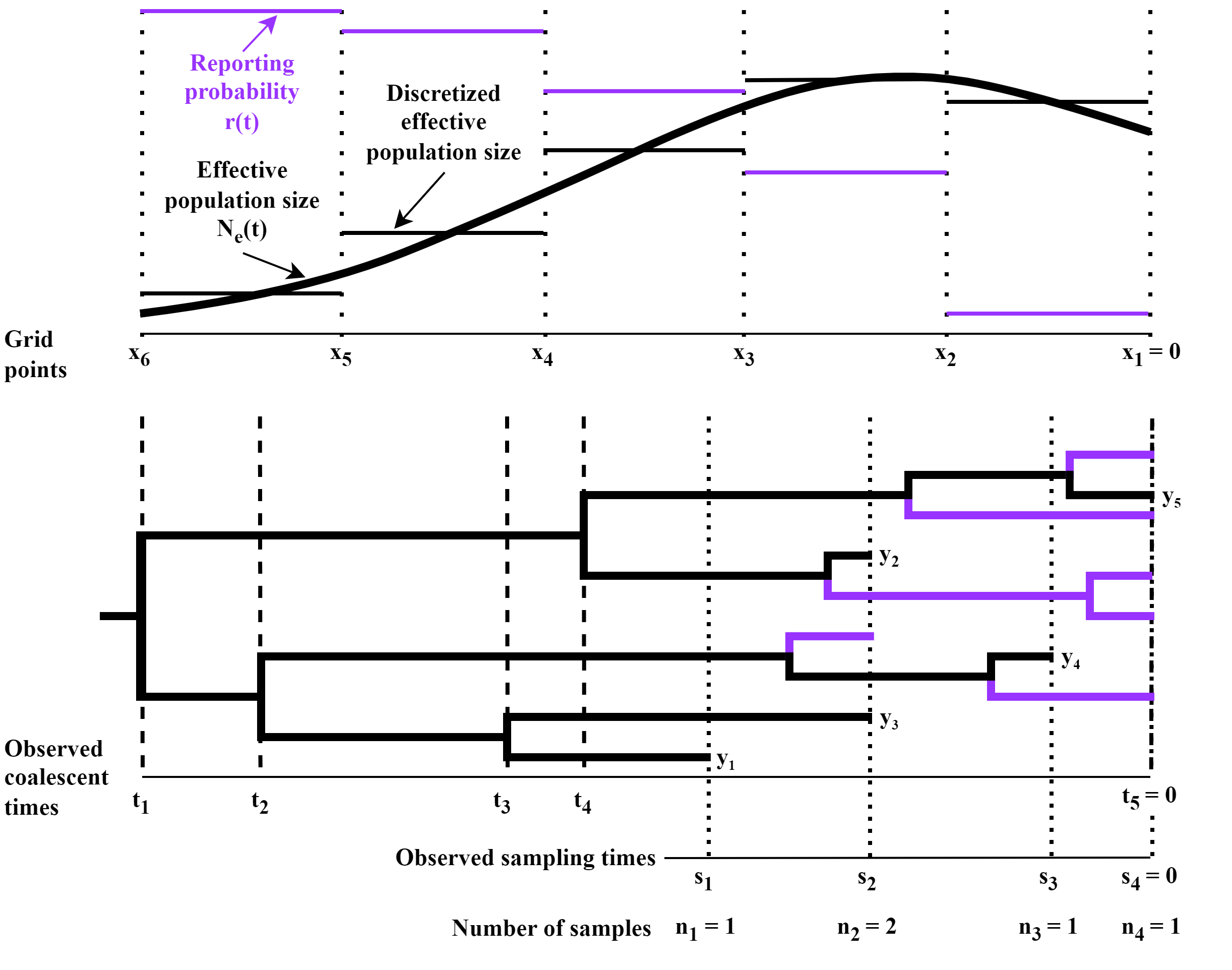}

}

\caption{\label{fig-genealogy-grid}Example of an effective population
size trajectory (top figure) and corresponding geneology (bottom
figure), with reporting probabilities and corresponding unobserved tips
denoted by purple coloring. For a real-time analysis the reporting
probability decreases as the collection date gets closer to present
time, time zero.}

\end{figure}%

When analyzing pathogen evolution, we use an alignment of sampled
pathogen genetic sequences as data. These sequences can either be
collected at the same time, isochronous sampling, or at different points
in time, heterochronous sampling. Here we are concerned with viruses
that evolve rapidly with continuously collected samples so we will
consider heterochronous sampling of DNA or RNA sequences aligned and
stored in matrix \(\boldsymbol{y} = \{y_{ji}\}\), \(j = 1, ... , n\),
\(i = 1, ..., L\), where \(n\) is the number of sequences and \(L\) is
the alignment length. The sequences, \(\boldsymbol{y}\), all ultimately
share a common ancestry, and the evolution of the sequences from their
most recent common ancestor is described by a bifurcating tree called a
genealogy, denoted as \(\boldsymbol{g}\).

We assume that given the genealogy, alignment sites are independent and
identically distributed. The evolutionary changes in the nucleotides
present at each alignment site, column of matrix \(\boldsymbol{y}\), are
modeled by a continuous-time Markov chain substitution model
parameterized by vector \(\boldsymbol{\theta}\). From a given viral
genealogy and substitution rate matrix, the probability of observing
sequences \(\boldsymbol{y}\),
\(P(\boldsymbol{y} | \boldsymbol{g}, \boldsymbol{\theta})\) can be
calculated using an efficient dynamic programming algorithm
\citep{felsenstein1981}. Equipped with a model for the alignment, a
model is needed for the pathogen's genealogy.

The lower half of Fig~\ref{fig-genealogy-grid} displays a genealogy
relating five sequences, black tree tips, collected across four sampling
times. Note the purple tips denote samples collected but not yet
reported and available for use by the time of analysis. Sampling times
are denoted by \(\boldsymbol{s} = \{s_j\}_{j=1}^{m}\) and sample sizes
by \(\boldsymbol{n} = \{n_j\}_{j=1}^{m}\) with
\(n = \sum_{j=1}^{m} n_j\). In this set up we imagine generating the
genealogy backwards in time starting from the most recent sampling time,
\(s_m = 0\). The branches of this evolutionary tree end at the sampling
times \(\boldsymbol{s}\), and the convergence, or coalescence, of two
branches corresponds to a common ancestor of the two sequences. The
tree's branches coalesce until the most recent common ancestor of all of
the samples, the root of the tree. The times of the coalescent events
are denoted \(\{t_i\}_{i=1}^{n-1}\), with \(t_1 > ... > t_{n-1}\), and
the last sampling time determines the last coalescent time,
\(t_n = s_m = 0\).

The effective population size, denoted \(N_e(t)\), is a time-varying
measure of genetic diversity. The number of active lineages at a time
\(t\) is the difference between the number of sampling and coalescent
events between times 0 and \(t\). The intervals \(I_{i,k}\) are defined
by the sampling and coalescent times, so the number of active lineages,
denoted \(l_{i, k}\), in an interval is constant. For \(k = 2, ..., n\),
the intervals that end in a coalescent event are denoted
\(I_{0, k} = (max\{t_k, s_j\}, t_{k - 1}]\), for \(s_j < t_{k - 1}\),
and intervals that end with a sampling event are denoted
\(I_{i, k} = (max\{t_k, s_{j + 1}\}, s_{j + i - 1}]\), for
\(t_k < s_{j + i - 1} \leq s_j < t_{k - 1}\) with \(i > 0\).

Coalescent models are continuous-time Markov chains used to model a
genealogy from a sample of sequences \citep{kingman1982}. Rodrigo et
al.~(1999) extended coalescent theory for heterochronous sampling to
calculate the joint distribution of a genealogy given its sampling
times, number of samples collected at each time, and the effective
population size, as the product of conditional densities and tail
probabilities of coalescent times \citep{rodrigo1999}:
\begin{equation}\phantomsection\label{eq-coalescent-density}{
\begin{split}
P(\boldsymbol{g} | \boldsymbol{s}, \boldsymbol{n}, N_e(t)) &= \prod_{k = 2}^{n}P(t_{k-1} | t_k, \boldsymbol{s}, N_e(t))  \\
&= \prod_{k = 2}^{n} \frac{A_{0, k}}{N_e(t_{k-1})}\exp \biggl\{  -\int_{I_{0, k}} \frac{A_{0, k}}{N_e(t)}dt - \sum\limits_{i \geq 1} \int_{I_{i,k}} \frac{A_{i, k}}{N_e(t)} dt \biggr\},
\end{split}
}\end{equation} with the coalescent factors
\(A_{i, k} = \binom{l_{i, k}}{2}\).

Assuming the effective population size trajectory \(N_e(t)\) is an
unknown function in continuous time, the integral in
Eq~\ref{eq-coalescent-density} is intractable. We adopt a common
approach \citep{palacios2012, gill2013, faulkner2020}, well described by
Lan et al.~(2015), that discretizes the effective population size to be
piecewise constant on a regular grid,
\(\boldsymbol{x} = \{x_d\}_{d=1}^{D}\), spanning from the most recent
sampling time, \(s_m = x_1\), to the first coalescent time,
\(t_1 = x_D\) \citep{lan2015}. In this approach we define
\(N_e(t) = \exp[\gamma(t)]\), and approximate \(N_e(t)\) by
\(N_e^{\gamma}(t) = \sum_{d = 1}^{D - 1} \exp(\gamma_d) 1_{t \in (x_d, x_{d + 1}]}\).
The \(\gamma_d\)'s \textit{a priori} follow a first order random walk:
\(\gamma_d | \gamma_{d - 1} \sim N(\gamma_{d - 1}, 1 / \kappa)\) with
\(\gamma_1 \sim N(0, \sigma_{\gamma}^2)\). We adopt the common approach
of using a gamma prior distribution for the hyperparameter \(\kappa\).

With heterochronous sampling, it is likely that the frequency of
sampling is related to the number of infections in the population (e.g.,
increased sampling intensity when there is an increase in infections).
Additional factors may also influence the sampling intensity, such as
time variable cost of sequencing a pathogen genome. In the preferential
sampling model, we model sampling events as a Poisson Process with
intensity \(\lambda(t)\) that depends on such time-varying factors:
\begin{equation}\phantomsection\label{eq-old-sampling-intensity}{
\begin{split}
    \log\lambda(t) =& \beta_0 + \beta_1 \log[N_e(t)] + \beta_2 f_2(t) + ... + \beta_p f_p(t),
\end{split}
}\end{equation} where \(f_2,...,f_m(t)\) are the additional time-varying
covariates \citep{karcher2020}. Note the sampling intensity can include
interactions between the covariates and the log effective population
size, but we do not include them in our model here. The coefficients
\(\boldsymbol{\beta} = (\beta_0, \beta_1, ..., \beta_p)\)'s are assigned
independent normal priors with means \(\boldsymbol{\mu}_{\beta}\) and
variances \(\boldsymbol{\sigma}_{\beta}\). Since the effective
population size is piecewise constant on the regular grid
\(\boldsymbol{x}\), for simplicity we require time-varying covariates
also be piecewise constant on the same grid.

Altogether, the posterior we are interested in is

\begin{equation}\phantomsection\label{eq-full-posterior}{
\begin{split}
    Pr(\boldsymbol{g}, \boldsymbol{\gamma}, \kappa, \boldsymbol{\beta}, \boldsymbol{\theta} | \boldsymbol{y}, \boldsymbol{s}) \propto & Pr(\boldsymbol{y} | \boldsymbol{g}, \boldsymbol{\theta}) Pr(\boldsymbol{g} | \boldsymbol{\gamma}, \boldsymbol{s}) Pr(\boldsymbol{s} | \boldsymbol{\gamma}, \boldsymbol{\beta}) Pr(\boldsymbol{\gamma} | \kappa)\\
    &Pr(\boldsymbol{\theta}) Pr(\kappa) Pr(\boldsymbol{\beta}).
\end{split}
}\end{equation} Approximation of this posterior via Markov Chain Monte
Carlo (MCMC) is implemented in the phylodynamic software BEAST
\citep{suchard2018, karcher2020}. This Bayesian inference is time and
memory intensive though, so it is common in practice to estimate the
genealogy first and assume the genealogy is known. When the genealogy is
known the posterior of interest reduces to

\begin{equation}\phantomsection\label{eq-reduced-posterior}{
  Pr(\boldsymbol{\gamma}, \kappa, \boldsymbol{\beta} | \boldsymbol{g}, \boldsymbol{s}) \propto Pr(\boldsymbol{g} | \boldsymbol{\gamma}, \boldsymbol{s}) Pr(\boldsymbol{s} | \boldsymbol{\gamma}, \boldsymbol{\beta}) Pr(\boldsymbol{\gamma} | \kappa) Pr(\kappa) Pr(\boldsymbol{\beta}).
}\end{equation} Approximations of this posterior via MCMC and via
Integrated Nested Laplace Approximations (INLA) \citep{palacios2012} are
implemented in the phylodynamic R package \texttt{phylodyn}
\citep{karcher2017}.

\subsection{Accounting for reporting
delays}\label{accounting-for-reporting-delays}

The time delay between collecting a sample and depositing that sample's
sequence into a database arose as a problem during the SARS-CoV-2
pandemic, because of the urgent need for up-to-date understanding of
disease dynamics. Missing the most recent data is especially problematic
for the preferential sampling model because of the dependency between
the sampling intensity and the effective population size. Intuitively, a
model that takes into account preferential sampling would underestimate
the effective population size close to the present time due to the lack
of observed samples. One possible solution of this problem is to use a
coalescent model without the preferential sampling component, avoiding
the dependency between the sampling intensity and the effective
population size. While the biases from the missing data would be avoided
with this strategy, unaccounted preferential sampling can result in
biases, and wider credible intervals than those modeled with
preferential sampling \citep{karcher2016}.

Another way to circumvent this missing data issue is to only use data up
to a time when all of the data is likely to have been reported (e.g.,
data up to two months prior to time of analysis). For example,
phylodynamics was used to compare SARS-CoV-2 lineages in England with
data truncated by two weeks to avoid reporting delays in 2021
\citep{volz2021}. The major pitfall of this truncation strategy is the
inability to perform real-time phylodynamics to inform outbreak
mitigation, a problem that increases for locations or time periods with
extensive reporting delays.

\subsection{Incorporating reporting delay distribution into preferential
sampling
model}\label{incorporating-reporting-delay-distribution-into-preferential-sampling-model}

To mitigate effects of reporting delays on real-time phylodynamic
analyses with preferential sampling, we propose incorporating
information about the distribution of recent delays in the sampling
intensity model. In the preferential sampling model sampling times are
modeled as a Poisson process with intensity \(\lambda(t)\). Let \(r(t)\)
be the probability that a sample collected at time \(t\) was sequenced
and reported by the time of the analysis. Define the observed sampling
times, \(\tilde{\boldsymbol{s}}\) to be the subset of the true sampling
times, \(\boldsymbol{s}\), that are reported by the time of analysis.
Then the observed sampling intensity, \(\tilde{\lambda}(t)\), could be
expressed as the product of the true sampling intensity and the
probability of a sample being reported, resulting in a thinned Poisson
process with intensity \(\tilde{\lambda}(t) = \lambda(t) r(t)\).
Plugging Eq~\ref{eq-old-sampling-intensity} into the definition of
\(\tilde{\lambda}(t)\), we get the following new model for the
log-sampling intensity

\begin{equation}\phantomsection\label{eq-new-sampling-intensity}{
  \log\tilde{\lambda}(t) = \log[r(t)] + \beta_0 + \beta_1 \log[N_e(t)] + \beta_2 f_2(t) + ... + \beta_p f_p(t).
}\end{equation}

\subsection{Implementation of reporting delay aware preferential
sampling
model}\label{implementation-of-reporting-delay-aware-preferential-sampling-model}

We developed a new version of \textbf{phylodyn} \citep{karcher2017},
\textbf{phylodyn2} (\url{https://github.com/CatalinaMedina/phylodyn2}),
which has a well-documented subset of the functionality of phylodyn,
with the additional ability to account for reporting delays in real-time
analyses by implementing our proposed reporting delay aware preferential
sampling model. In this implementation, the reporting probabilities are
assumed to be known. The empirical cumulative distribution of recent
reporting delays is used to calculate the reporting probabilities,
\(\boldsymbol{r} = (r_1, r_2, ..., r_D)\). Similarly to the effective
population size and any covariates, the reporting probabilities
\(\boldsymbol{r}\) are also defined as piecewise constant across the
regular grid \(\boldsymbol{x}\).

The R package phylodyn included several posterior sampling strategies.
For phylodyn2 we chose to focus on the INLA based strategy to
approximate the marginal posterior distributions
\(Pr(\gamma_i | \boldsymbol{g})\). This INLA implementation formulates
the model for sampling times as a Poisson regression. However, the
original phylodyn implementation did not allow for inclusion of a user
specified offset term into this regression. In phylodyn2 we added an
offset term to the original Poisson regression, where this offset term
is set to \(log[r(t)]\), calculated from a user specified vector of
recent reporting delays.

One could view the \(log[r(t)]\) term as a time-varying covariate of the
sampling intensity, with coefficient one. The appeal of this perspective
is the ease of use for phylodynamic tools that allow for time-varying
covariates in the sampling intensity, such as BEAST and phylodyn. One
could specify \(log[r(t)]\) as a regression covariate with a narrow
prior for the coefficient of this term centered at one. This adds
unnecessary randomness, since the coefficient of this term is
theoretically one, but the ease of use makes this option worth
exploring. This implementation is also available in phylodyn2, and its
performance is examined in the supplementary materials.

All code to reproduce the results in this paper can be found at
\url{https://github.com/CatalinaMedina/reporting-delays-in-phylodynamics-paper}.

\section{Results}\label{results}

\subsection{Simulations}\label{simulations}

We performed simulation studies to mimic real-time phylodynamic analyses
in the presence of preferential sampling, aiming at two primary
objectives. Firstly, to investigate the effects of reporting delays with
currently available phylodynamic inferential strategies. Secondly, to
compare the performance of our proposed model against the currently
available strategies. Of key interest is how well the effective
population size trajectory can be inferred close to the most recent
sampling time.

Three real-time inferential strategies were considered for comparison:
avoid modeling the sampling time dependency by using the the Bayesian
nonparametric phylodynamic reconstruction (BNPR) model, model the
sampling time dependency with the Bayesian nonparametric phylodynamic
reconstruction with preferential sampling (BNPR PS) model, and model the
sampling time dependency and reporting delays with our proposed
reporting delay aware preferential sampling model. We also fit the BNPR
PS model to all of the data, regardless of whether it was reported, to
provide a retrospective baseline for the performance of these real-time
inferential strategies.

\begin{figure}

\centering{

\includegraphics[width=\textwidth]{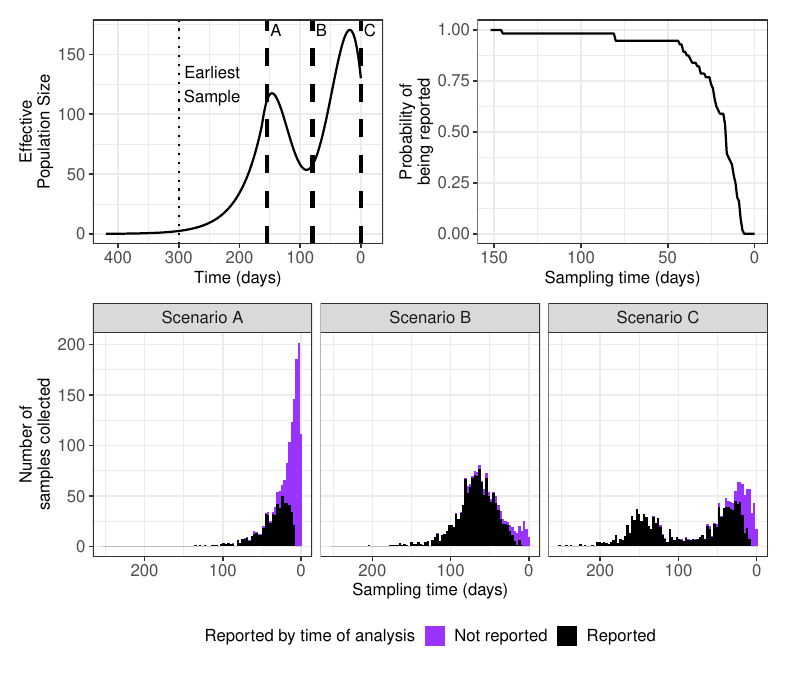}

}

\caption{\label{fig-simulation-set-up}Simulation details: effective
population trajectories (upper left plot), reporting probability by
sampling time (upper right plot) obtained from the Washington state
data, and histograms of sampling times from the last simulation of in
each simulation scenario, approxiamtely 1500 samples each, colored by
whether sample was reported by time of analysis (bottom plots). Each
simulation scenario had a different time zero, i.e., time of latest
sample (dashed lines). The earliest sampling time in each scenario was
at the same point in the trajectory (dotted line).}

\end{figure}

We used three simulation scenarios with the same effective population
size trajectory, but across different time periods so that the effects
of reporting delays with different trajectory behavior near time zero
could be investigated. The upper-left panel of
Fig~\ref{fig-simulation-set-up} shows the effective population size
trajectory, as well as the most recent sampling time, the time of
analysis, for each scenario. Since time is viewed in reverse, the most
recent sample in simulation scenario C is time zero, and the earliest
sample was 300 days prior. Scenarios A and B had sampling time periods
of 150 days and 220 days, respectively. Scenario A is meant to resemble
an initial outbreak, which would have fewest samples due to reporting
delays. Scenario B allows us to examine behavior when there is an
increase occurring near present time, but less sampling. Lastly, in
scenario C there is a decline near present time and the recent peak
corresponds to more reported samples near time zero than scenario B.

Sampling times were simulated from an inhomogeneous Poisson process with
intensity \(\lambda(t) = \exp(\beta_0) [N_e(t)]^{\beta_1}\). Coalescent
times were simulated using a time-transformation technique where the
coalescent likelihood is treated as an inhomogeneous Poisson process
\citep{slatkin1991}. Parameter \(\beta_1\) was set to 2 to create a
reasonably strong preferential sampling effect and \(\exp(\beta_0)\) was
selected to achieve a sample size of approximately 1500 samples, each
with its own sampling time.

For each sampling time we simulated a random Bernoulli to indicate if a
sample was reported by the time of analysis. Sampling times for each
scenario are plotted in Fig~\ref{fig-simulation-set-up} and colored by
whether it was observed or not. To create realistic delays, the
reporting probabilities were obtained from the empirical reporting delay
distribution of SARS-CoV-2 sequences collected in the state of
Washington. See the real data investigation results subsection for
details, visualized in upper-right panel of
Fig~\ref{fig-simulation-set-up}. The tips of the genealogy of the full
tree that correspond to unreported samples were pruned from the tree, to
get the observed genealogy. Each inference was performed with the
INLA-based Bayesian phylodynamic inference implemented in the R package
phylodyn2.

\begin{figure}

\centering{

\includegraphics[width=\textwidth]{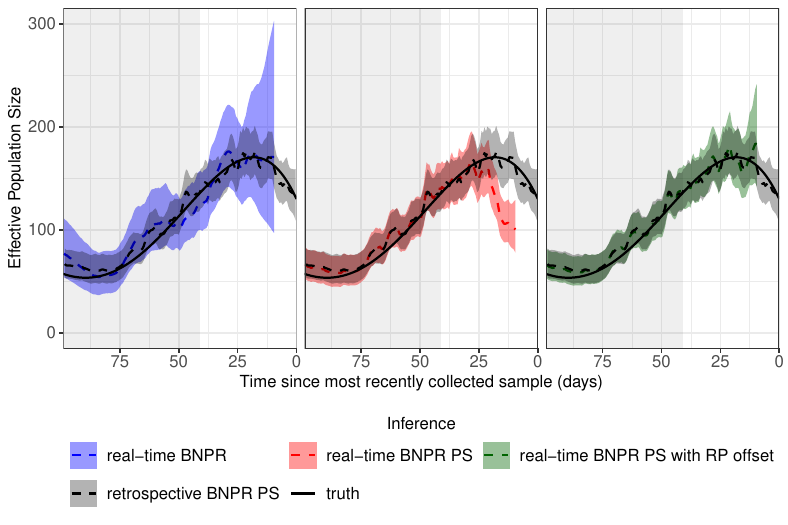}

}

\caption{\label{fig-sim-scenario-c-comparison}Comparison of real-time
phylodynamic methods to infer the effective population size trajectory
for a single simulation in scenario C, with reporting delays and
preferential sampling present. Median estimates of the effective
population size and 95\% credible intervals are plotted. The white
background indicates the recent time period likely suffering from
reporting delays, specifically where reporting probabilities (RPs) are
below 90\%, and is therefore the region of interest.}

\end{figure}

We will begin by discussing the results of a single simulation within a
scenario, in order to better understand the patterns in the performance
of each inference strategy across all of the simulations.
Fig~\ref{fig-sim-scenario-c-comparison} plots the true (solid lines) and
inferred (dashed lines) effective population size trajectory for the 100
days prior to the most recently collected sample. Here we focus on the
two options of real-time inference, the BNPR and BNPR PS models, and our
proposed inclusion of reporting probabilities in the BNPR PS model.
While the ultimate goal is to be able to infer the true effective
population size trajectory, it is useful to see how closely the data
generating model can approximate the true trajectory, within simulation
scenario C, where the trajectory of interest is on the decline at
present time. This is why each plot also contains the BNPR PS inference
performed retrospectively on all of the data, not just the observed data
-- this serves as a baseline to compare the inference of \(N_e(t)\) from
each real-time inferential method. The white background indicates the
time period of interest, where delays are probable, and conversely the
gray background indicates the period where reporting delays are
unlikely. We chose to use the 90th percentile of the Washington state
data reporting delays distribution, which was 41 days in this case, as
the cutoff for these two periods.

In the first panel of Fig~\ref{fig-sim-scenario-c-comparison} we see
real-time inference with the BNPR model, which ignores the dependency
between \(N_e(t)\) and the sampling time. The BNPR model appears to have
relatively low bias, but wide 95\% credible intervals that increase in
width near time zero. The real-time inference with the BNPR PS model
stands out because of the bias which increases as time approaches the
most recently collected sample. This demonstrates the bias introduced
when using the data generating model, the BNPR PS model, when there are
reporting delays present in the data. Alternatively, our implementation
of the BNPR PS model with the reporting probabilities included as an
offset in the sampling intensity, has less bias than the BNPR PS model
near time zero and visibly narrower 95\% credible intervals than the
BNPR model near time zero.

\begin{figure}

\centering{

\includegraphics[width=\textwidth]{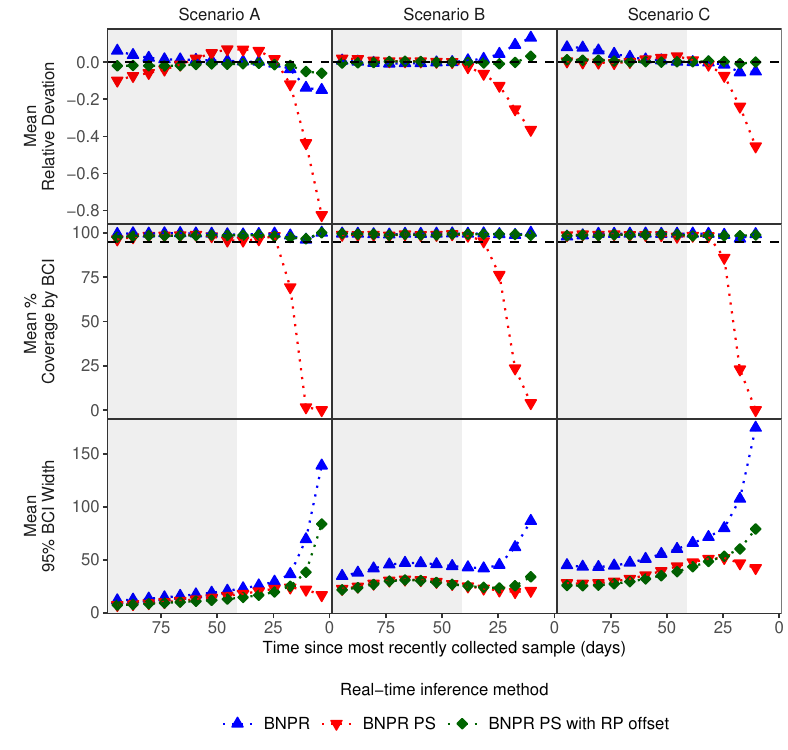}

}

\caption{\label{fig-sim-estimation-evaluation}Seven-day moving averages
of the mean relative deviation, mean percent coverage, and mean interval
width by the 95\% Bayesian credible intervals, for each real-time
phylodynamic strategies to infer the effective population size in each
simulation scenario with preferential sampling (PS) and reporting delays
in the observed data. Inference was performed with Bayesian
nonparametric phylodynamic reconstruction (BNPR), BNPR PS, and with BNPR
PS with reporting probabilities (RP) in sampling intensity as an
offset.}

\end{figure}%

The results identified from the single simulation in
Fig~\ref{fig-sim-scenario-c-comparison} generally persist across all 500
simulations, in each of the three simulation scenarios, visualized in
Fig~\ref{fig-sim-estimation-evaluation}. The plots present a seven-day
moving average of the mean relative deviation, mean percent of 95\%
Bayesian credible intervals which covered the true value, and mean 95\%
credible interval width for each inference strategy in each simulation
scenario. A moving average was chosen because a metric of the inference
over the entire time period would be insufficient to describe how
inference performance changes with proximity to time zero. Mean relative
deviation is the most important of the three chosen metrics because it
assess accuracy, of the point estimate, interval coverage was selected
to examine the accuracy of the uncertainty of the estimates, and
interval width was useful for assessing precision to compare those
models with good accuracy and good coverage. Since the performance of
these estimation strategies near time zero is of key interest, these
plots were truncated to the most recent 100 days. To view the
performance metrics results for all inferential methods considered see
the supplemental material Table S1, Table S2, and Table S3.

Focusing on the time period of interest, the most recent 41 days, our
proposed reporting delays aware BNPR PS model consistently has lower
mean relative deviation than the BNPR method, in each seven-day moving
average, though there is not much practical difference. The BNPR PS
model has increasing relatively large mean relative deviation as
sampling times decrease to time zero, in each simulation scenario. The
absolute maximum mean relative deviations in scenario A are all achieved
in the week prior to time zero are 0.15, 0.82, and 0.06 for the BNPR,
BNPR PS, and our reporting delay aware model respectively. The 95\%
Bayesian credible intervals for the BNPR and our reporting delays aware
BNPR PS model are consistently conservative, while the BNPR PS model's
95\% credible intervals' coverage drops below 95\% and approaches 0\% as
sampling time approaches time zero. Finally, while maintaining
competitively low bias and high coverage, our proposed model
consistently has lower mean 95\% Bayesian credible interval widths than
the BNPR model, with the difference between the two models increasing as
sampling time approaches time zero.

\subsection{Real data investigation: Washington state COVID
dynamics}\label{real-data-investigation-washington-state-covid-dynamics}

\begin{figure}

\centering{

\includegraphics[width=\textwidth]{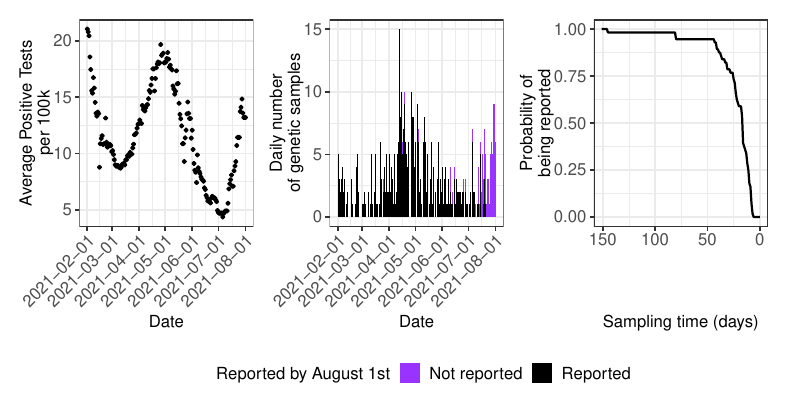}

}

\caption{\label{fig-real-data-with-delays}Left plot is the average
number of positive COVID-19 tests for most recent seven days per 100,000
people in the state of Washington. Middle panel shows number of
SARS-CoV-2 genetic samples collected in Washington state, colored by
whether the sample was reported by the time of analysis, August 1, 2021.
Right panel shows empirical cumulative distribution of reporting delays
from the month prior to time of analysis.}

\end{figure}

We used SARS-CoV-2 sequences from Washington state for the purpose of
investigating the differences between a real-time phylodynamic analysis
with and without our proposed method to account for reporting delays in
genomic data. The SARS-CoV-2 sequences were accessed via the GISAID
database available at \url{https://gisaid.org/EPI_SET_220330me}, for
Washington state sampled between February 01, 2021 and August 01, 2021,
inclusive \citep{shu2017}. This time period was of interest because
researchers were regularly sequencing Washington samples at this point
in the pandemic, and the reporting behavior is relatively consistent
during this period. Fig~\ref{fig-real-data-with-delays} plots seven day
averages of the number of COVID-19 cases per 100,000 people in
population in the state of Washington, the daily number of SARS-CoV-2
samples available in GISAID for Washington, colored by whether the
sequence was sampled by August 01, 2021 (left plot), and the empirical
cumulative distribution function for sampling dates between July 01,
2021 and August 01, 2021.

The observed data are samples that had been sequenced and reported to
GISAID on or before August 01, 2021, time zero of our analysis. The 90th
percentile of the reporting delays distribution is 41 days. Since we are
interested in the inference of \(N_e(t)\) when reporting delays are
present, we chose to focus our attention on the most recent 41 days.

Genealogy estimation was performed in BEAST for each data set: the full
500 sequences, observed 412 sequences, and 375 remaining sequences after
truncation. We used the HKY substitution model with empirically
estimated base frequencies \citep{hasegawa1985}, Bayesian Skygrid
coalescent model \citep{gill2013, drummond2002}, and a Uniform prior on
the clock rate between \(2.38\times10^{-3}\) and \(8\times10^{-4}\)
\citep{neher2022}. The MCMC was run for \(25 \times 10^6\) iterations,
logging parameters every 2000th iteration. The maximum clade credibility
tree of the posteriors were used as the known genealogy in the
phylodynamic reconstruction, for each of the three analyses. See the
supplemental materials section for more details about this analysis to
obtain the genealogies.

\begin{figure}

\centering{

\includegraphics[width=\textwidth]{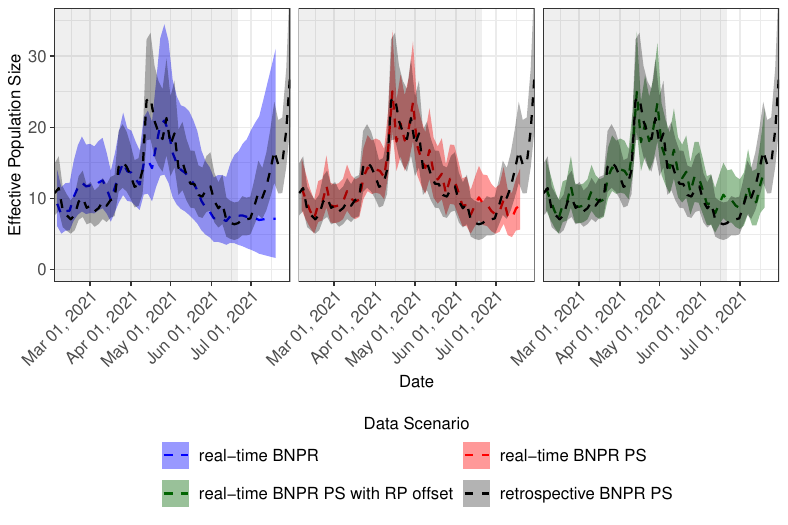}

}

\caption{\label{fig-Washington-inference-comparison}Bayesian
nonparametric phylodynamic reconstruction (BNPR) methods used to infer
effective population size trajectory for SARS-CoV-19 in Washington
state. Each panel shows the inference from a real-time analysis on data
suffering from reporting delays and from a retrospective analysis with
completely reported data. The white background indicates the recent time
period likely suffering from reporting delays.}

\end{figure}

Inference of the effective population size was performed with the same
strategies used in our simulations to compare the performance of our
proposed methods against available options.
Fig~\ref{fig-Washington-inference-comparison} shows the inference of the
effective population size for three modeling strategy: Bayesian
nonparametric phylodynamic reconstruction (BNPR), BNPR with preferential
sampling (BNPR PS), and our proposed inclusion of reporting
probabilities in the BNPR PS model as an offset.

The retrospective analysis with all of the collected samples with the
BNPR PS model infers a peak in transmission activity in mid April of
2021, dropping to a minimum in mid June, followed by a steady increase
continuing into August 2021. The results for these analyses are
consistent with the trajectory of COVID-19 cases for this time period,
visualized in Fig~\ref{fig-real-data-with-delays}, with approximately a
two week delay which could be due to reporting delays in COVID test
results.

When comparing the real-time analyses we see similar patterns as those
identified in our simulations. Using the retrospective BNPR PS model for
comparison we see near real time, the BNPR model suffers from low
precision, the BNPR PS model's credible intervals disagree with the
retrospective analysis credible intervals, and our proposed BNPR PS with
the reporting probability correction is consistent with the
retrospective BNPR PS model results, with higher precision than the BNPR
model. This gain in precision found with our proposed model would have
allowed real-time analysis to infer the increase near present time that
the two currently available competitive methods underestimated.

\section{Discussion}\label{discussion}

In this work we investigated the effects of reporting delays on
real-time phylodynamic methods to infer the effective population size
and we proposed a reporting delay aware model. Through simulations we
demonstrated that when preferential sampling is present, real-time
analysis with the BNPR PS model suffers from increasingly extreme bias
when inferring the effective population size near present time, implying
that the BNPR PS model should not be used for such cases because it is
unreliable. We also showed that when the preferential sampling
relationship is not modeled, real-time analysis with the BNPR model has
less bias than the BNPR PS model, but is largely uninformative due to
its low precision near present time. Across simulations we found our
reporting delay aware model to performed comparably to the BNPR model,
not suffering from the same biases as the BNPR PS model, and with
increased data our model obtains increased precision near present time,
relative to the BNPR model. Our results support the intuition that we
can infer \(N_e(t)\) more accurately and precisely with more data,
specifically when there are more samples sequenced and available for
analysis. Beyond the simulations, our Washington data analysis found
evidence of preferential sampling and behavior consistent with out
simulation results: we saw agreeable results between our reporting delay
aware model and the retrospective BNPR PS model, the real-time BNPR
model had very low precision near present time, and the real-time BNPR
PS model strongly disagreed with the retrospective BNPR PS model. The
simulated and real data results provide compelling arguments that
reporting delays should not be ignored in real-time analysis, and that
the effective population size trajectory is a reasonable indicator for
the effective number of infections.

For simplicity, we assumed the pathogen genealogy is known in our
implementation of the reporting delay aware model in phylodyn2,
obtaining the marginal posteriors of \(N_e(t)\) with INLA. The value of
this choice is that it is fast and can handle much larger number of
sequences than BEAST which jointly infers the genealogy and other model
parameters, including \(N_e(t)\). Computational speed and feasibility
are necessary considerations with Bayesian phylodynamic methods,
especially with online surveillance. A natural next step from this work
would be to incorporate our reporting probability adjustment into the
joint posterior inferred by the BNPR PS model in BEAST.

Our reporting delay aware model currently assumes that reporting
probabilities are known, and our implementation uses recent reporting
delays to estimate current reporting probabilities. This strategy is
limited to locations, times, lineages, and even laboratories where there
is believed to be consistency in reporting delays for sequences
\citep{petrone2023}. As such, care is necessary when defining the
reporting probability distribution for use in the sampling intensity of
our model. The next extension of this work would be to jointly infer the
reporting probabilities and the effective population size. This could
allow for increased accuracy and better reflect our uncertainty about
reporting probabilities, especially for areas with rapid changes in
reporting behavior. Perhaps of most interest would be to allow for the
reporting delay distribution to change overtime, allowing for updated
surveillance of the effective population size with continual data
collection.

The BNPR PS method models the sampling intensity parametrically, so
naturally there may be concern of model misspecification, especially
when studying new variants of unknown infectiousness. Cappello and
Palacios (2022) proposed a model which allows for the relationship
between the effective population size and the sampling intensity to vary
with time as follows: \(\lambda(t) = \beta(t) N_e(t)\), where
\(\beta(t)\) is inferred nonparametrically from the genetic and sampling
time data \citep{cappello2022}. It would be of interest to extend this
model to incorporate known reporting probabilities. The next question
would be if it could jointly infer reporting probabilities and
\(N_e(t)\) with the time-varying \(\beta(t)\).

Our proposed reporting delay aware model is a first step in mitigating
the effects of reporting delays on real-time phylodynamic analyses. This
work has important implications for real-time research with genomic
data. We identified that the data generating model can be biased when
ignoring the presence of missing data near present time due to reporting
delays. The severity of this bias increases as the number of sequences
observed decreases, but this bias can be corrected by using historical
data about reporting delays.

\section{Acknowledgements}\label{acknowledgements}

We gratefully acknowledge all data contributors, i.e., the authors and
their originating laboratories responsible for obtaining the specimens,
and their submitting laboratories for generating the genetic sequence
and metadata and sharing via the GISAID Initiative, on which this
research is based.

This work was in part funded by the UC Irvine Investing to Develop
Center-Scale Multidisciplinary Convergence Research Programs Seed
Funding Award and by the UC CDPH Modeling Consortium. Julia Palacios
acknowledges support from NSF grant DMS-2143242, NIH grant
R35GM14833801. We thank Lorenzo Cappello for useful discussions during
the early stages of this project.

\bibliographystyle{biom}  
\bibliography{references.bib}

\clearpage

\appendix

\setcounter{table}{0}
\renewcommand{\thetable}{S-\arabic{table}}
\renewcommand{\thefigure}{S-\arabic{figure}}
\renewcommand{\thesection}{S-\arabic{section}}
\setcounter{equation}{0}
\setcounter{section}{0}
\setcounter{figure}{0}

\begin{center}
\Large \bf
Supplementary Materials
\end{center}

\section{Simulation results}\label{simulation-results}

In the manuscript we focused on three real-time inferential strategies:
BNPR, BNPR PS, and our proposed reporting delay aware BNPR PS model with
reporting probabilities incorporated as an offset in the model of the
sampling intensity. For completeness, we also considered the truncation
technique, where only samples collected up until some point in time are
used. This is meant to avoid the most recent time period, and therefore
avoid reporting delays in data, but by definition is not much of a
real-time analysis. Regardless, we investigated this strategy to show
that the truncation technique can result in important recent behavior
remaining unknown.

We also considered an alternative implementation of our reporting delay
aware preferential sampling model. This alternative implementation is
discussed in the last subsection of our methods section in the
manuscript. The implementation with reporting probabilities being
incorporated as an offset is useful because any existing software that
can incorporate a time varying covariate into the log sampling intensity
equation would be automatically able to implement our proposed model. As
expected, this implementation performed similarly to our originally
proposed implementation, with the caveat of increased variation due to
the need to model the coefficient of this regression term, which is
known to be one in theory.

We also additionally consider the BNPR model used retrospectively on all
data. We included this because, while the BNPR PS model is the data
generating model in this case, it is less commonly used than the BNPR
model. An important finding is that the real-time inference with our
proposed reporting delay aware BNPR PS model on only reported samples is
competitive with the retrospective BNPR model fit to all of the data,
further supporting why our model should be used.

\subsection{Further simulation results with Washingon state's reporting
delays}\label{further-simulation-results-with-washgtinon-states-reporting-delays}

\subsubsection{Results from last simulation in each
scenario}\label{results-from-last-simulation-in-each-scenario}

\begin{figure}[H]

{\centering \includegraphics[width=\textwidth]{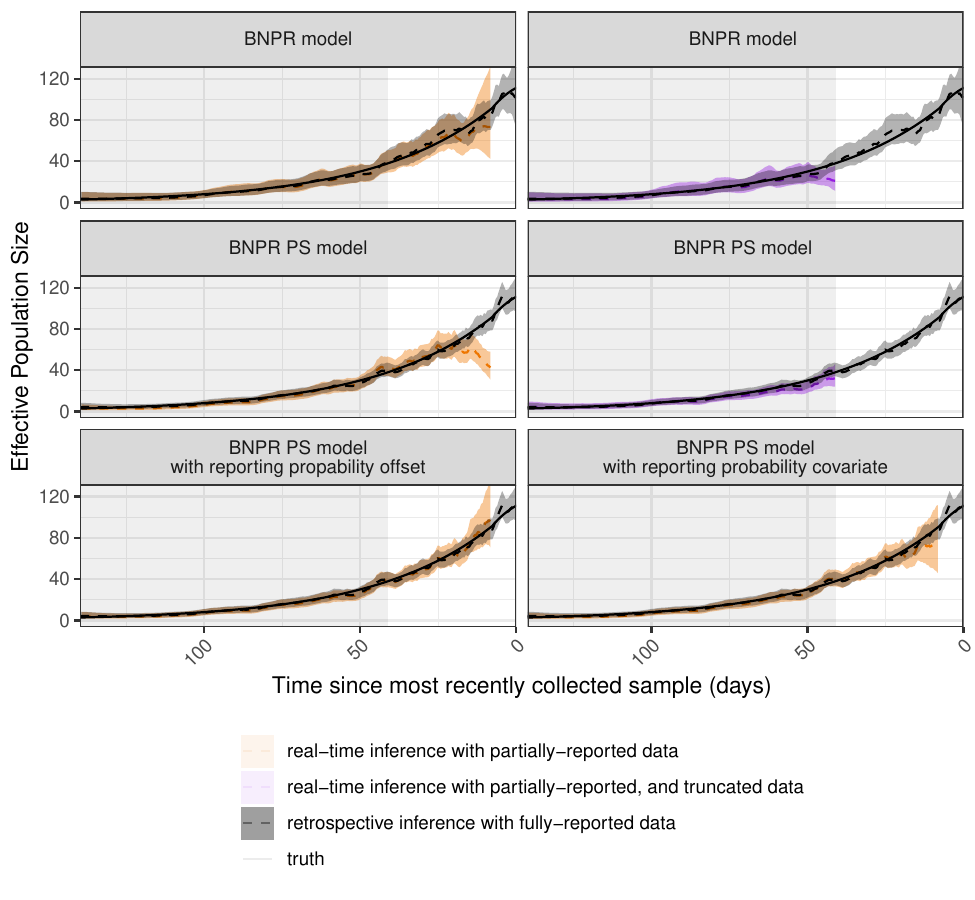}

}

\caption{Comparison of phylodynamic estimation methods of effective
population size trajectory for three different simulated data scenarios
from the scenario A trajectory.}

\end{figure}%

\begin{figure}[H]

{\centering \includegraphics[width=\textwidth]{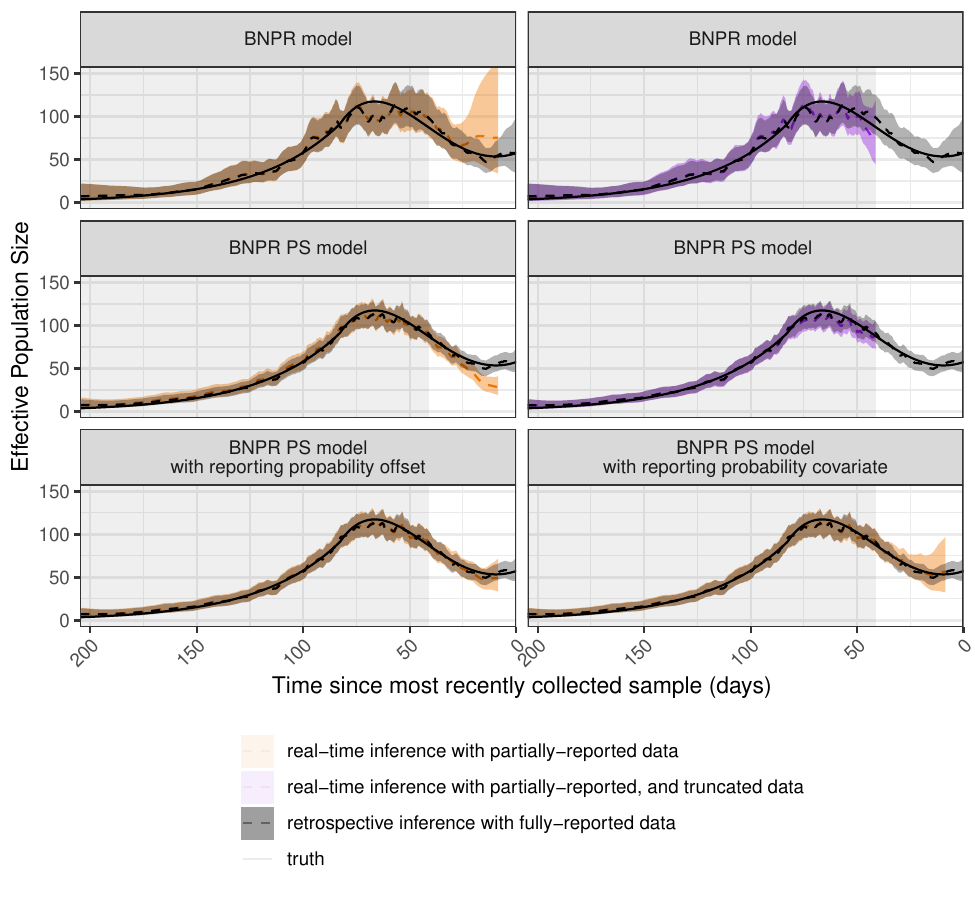}

}

\caption{Comparison of phylodynamic estimation methods of effective
population size trajectory for three different simulated data scenarios
from the scenario B trajectory.}

\end{figure}%

\begin{figure}[H]

{\centering \includegraphics[width=\textwidth]{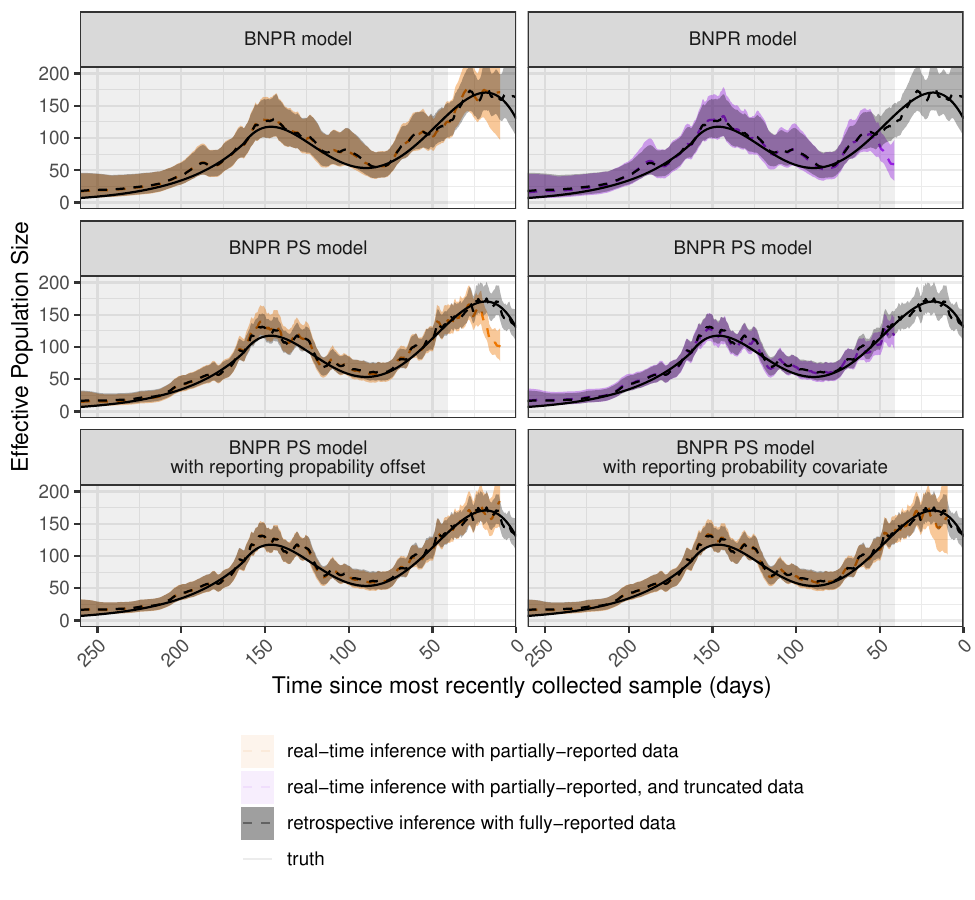}

}

\caption{Comparison of phylodynamic estimation methods of effective
population size trajectory for three different simulated data scenarios
from the scenario C trajectory.}

\end{figure}%

\subsubsection{Performance metrics across all
simulations}\label{performance-metrics-across-all-simulations}

\begin{figure}[H]

{\centering \includegraphics[width=\textwidth]{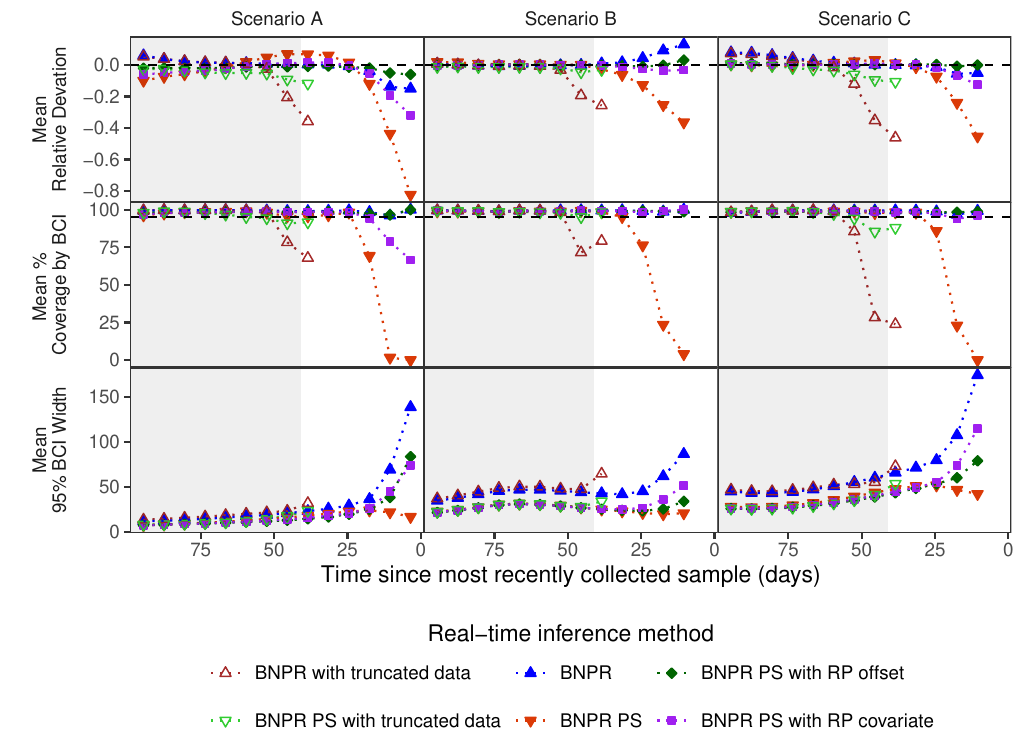}

}

\caption{Seven-day moving averages of the mean relative deviation, mean
percent coverage, and mean interval width by the 95\% Bayesian credible
intervals, for each real-time phylodynamic strategies to infer the
effective population size in each simulation scenario with preferential
sampling (PS) and reporting delays in the observed data. Inference was
performed with Bayesian nonparametric phylodynamic reconstruction
(BNPR), BNPR PS, and with BNPR PS with reporting probabilities (RP) in
sampling intensity as an offset.}

\end{figure}%

\begin{figure}[H]

{\centering \includegraphics[width=\textwidth]{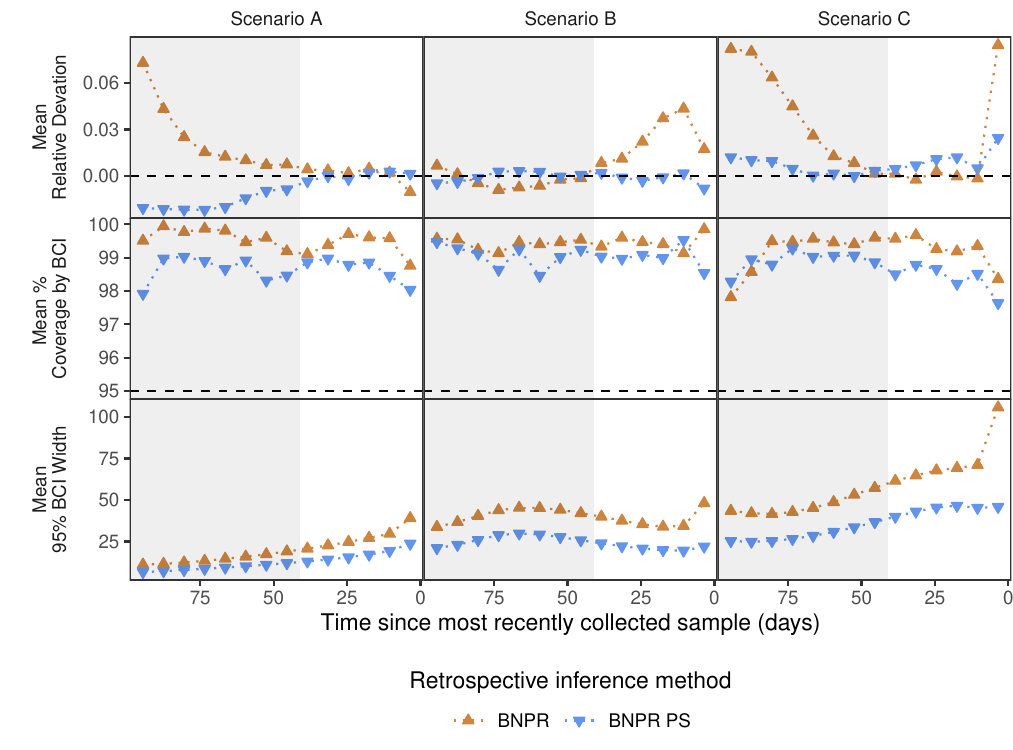}

}

\caption{Seven-day moving averages of the mean relative deviation, mean
percent coverage, and mean interval width by the 95\% Bayesian credible
intervals, for each retrospective phylodynamic strategy to infer the
effective population size in each simulation scenario with preferential
sampling (PS) and no reporting delays in the observed data. Inference
was performed with Bayesian nonparametric phylodynamic reconstruction
(BNPR) and BNPR PS.}

\end{figure}%

\newpage

\begingroup\fontsize{7}{9}\selectfont

\begin{longtable}[t]{>{\raggedright\arraybackslash}p{0.5in}>{\raggedleft\arraybackslash}p{0.325in}>{\raggedleft\arraybackslash}p{0.325in}>{\raggedleft\arraybackslash}p{0.325in}>{\raggedleft\arraybackslash}p{0.325in}>{\raggedleft\arraybackslash}p{0.325in}>{\raggedleft\arraybackslash}p{0.325in}>{\raggedleft\arraybackslash}p{0.325in}>{\raggedleft\arraybackslash}p{0.325in}}
\caption{Mean relative deviation of all inferential techniques for effective
population size in each data scenario for 500 simulations in scenarios
A, B, and C. Time periods span approximately from first simulated
sampling time to 90th percentil of historic reporting delays, 41 days.
The final time period for each simulation scenario spans from the first
to last sampling time simulated.}\tabularnewline

\toprule
\multicolumn{1}{c}{ } & \multicolumn{2}{c}{Retrospective} & \multicolumn{6}{c}{Real-time inference} \\
\cmidrule(l{3pt}r{3pt}){2-3} \cmidrule(l{3pt}r{3pt}){4-9}
Time period (days) & BNPR & BNPR PS & Trunc. BNPR & Tunc. BNPR PS & BNPR & BNPR PS & BNPR PS with RP offset & BNPR PS with RP covariate\\
\midrule
\addlinespace[0.3em]
\multicolumn{9}{l}{\textbf{Scenario A}}\\
\hspace{1em}\cellcolor{gray!10}{{}[0,7)} & \cellcolor{gray!10}{-0.01} & \cellcolor{gray!10}{0.00} & \cellcolor{gray!10}{} & \cellcolor{gray!10}{} & \cellcolor{gray!10}{-0.15} & \cellcolor{gray!10}{-0.82} & \cellcolor{gray!10}{-0.06} & \cellcolor{gray!10}{-0.32}\\
\hspace{1em}{}[7,14) & 0.00 & 0.00 &  &  & -0.14 & -0.44 & -0.05 & -0.19\\
\hspace{1em}\cellcolor{gray!10}{{}[14,21)} & \cellcolor{gray!10}{0.00} & \cellcolor{gray!10}{0.00} & \cellcolor{gray!10}{} & \cellcolor{gray!10}{} & \cellcolor{gray!10}{-0.04} & \cellcolor{gray!10}{-0.12} & \cellcolor{gray!10}{-0.02} & \cellcolor{gray!10}{-0.05}\\
\hspace{1em}{}[21,28) & 0.00 & 0.00 &  &  & -0.01 & 0.02 & -0.01 & 0.00\\
\hspace{1em}\cellcolor{gray!10}{{}[28,35)} & \cellcolor{gray!10}{0.00} & \cellcolor{gray!10}{0.00} & \cellcolor{gray!10}{} & \cellcolor{gray!10}{} & \cellcolor{gray!10}{0.00} & \cellcolor{gray!10}{0.06} & \cellcolor{gray!10}{-0.01} & \cellcolor{gray!10}{0.02}\\
\hspace{1em}{}[35,42) & 0.00 & 0.00 & -0.36 & -0.12 & 0.00 & 0.07 & -0.01 & 0.01\\
\hspace{1em}\cellcolor{gray!10}{{}[0,154]} & \cellcolor{gray!10}{0.12} & \cellcolor{gray!10}{0.05} & \cellcolor{gray!10}{0.10} & \cellcolor{gray!10}{0.05} & \cellcolor{gray!10}{0.10} & \cellcolor{gray!10}{-0.02} & \cellcolor{gray!10}{0.05} & \cellcolor{gray!10}{0.01}\\
\addlinespace[0.3em]
\multicolumn{9}{l}{\textbf{Scenario B}}\\
\hspace{1em}{}[7,14) & 0.04 & 0.00 &  &  & 0.13 & -0.36 & 0.03 & -0.03\\
\hspace{1em}\cellcolor{gray!10}{{}[14,21)} & \cellcolor{gray!10}{0.04} & \cellcolor{gray!10}{0.00} & \cellcolor{gray!10}{} & \cellcolor{gray!10}{} & \cellcolor{gray!10}{0.09} & \cellcolor{gray!10}{-0.25} & \cellcolor{gray!10}{0.00} & \cellcolor{gray!10}{-0.04}\\
\hspace{1em}{}[21,28) & 0.02 & 0.00 &  &  & 0.04 & -0.13 & -0.01 & -0.02\\
\hspace{1em}\cellcolor{gray!10}{{}[28,35)} & \cellcolor{gray!10}{0.01} & \cellcolor{gray!10}{0.00} & \cellcolor{gray!10}{} & \cellcolor{gray!10}{} & \cellcolor{gray!10}{0.02} & \cellcolor{gray!10}{-0.06} & \cellcolor{gray!10}{0.00} & \cellcolor{gray!10}{-0.01}\\
\hspace{1em}{}[35,42) & 0.01 & 0.00 & -0.26 & -0.03 & 0.01 & -0.03 & 0.00 & 0.00\\
\hspace{1em}\cellcolor{gray!10}{{}[0,228]} & \cellcolor{gray!10}{0.14} & \cellcolor{gray!10}{0.07} & \cellcolor{gray!10}{0.12} & \cellcolor{gray!10}{0.06} & \cellcolor{gray!10}{0.14} & \cellcolor{gray!10}{0.07} & \cellcolor{gray!10}{0.06} & \cellcolor{gray!10}{0.06}\\
\addlinespace[0.3em]
\multicolumn{9}{l}{\textbf{Scenario C}}\\
\hspace{1em}{}[7,14) & 0.00 & 0.01 &  &  & -0.05 & -0.45 & 0.00 & -0.13\\
\hspace{1em}\cellcolor{gray!10}{{}[14,21)} & \cellcolor{gray!10}{0.00} & \cellcolor{gray!10}{0.01} & \cellcolor{gray!10}{} & \cellcolor{gray!10}{} & \cellcolor{gray!10}{-0.06} & \cellcolor{gray!10}{-0.24} & \cellcolor{gray!10}{-0.01} & \cellcolor{gray!10}{-0.07}\\
\hspace{1em}{}[21,28) & 0.00 & 0.01 &  &  & -0.01 & -0.07 & 0.00 & -0.02\\
\hspace{1em}\cellcolor{gray!10}{{}[28,35)} & \cellcolor{gray!10}{0.00} & \cellcolor{gray!10}{0.01} & \cellcolor{gray!10}{} & \cellcolor{gray!10}{} & \cellcolor{gray!10}{0.00} & \cellcolor{gray!10}{-0.01} & \cellcolor{gray!10}{0.01} & \cellcolor{gray!10}{0.00}\\
\hspace{1em}{}[35,42) & 0.00 & 0.00 & -0.46 & -0.11 & 0.00 & 0.01 & 0.00 & 0.00\\
\hspace{1em}\cellcolor{gray!10}{{}[0,307]} & \cellcolor{gray!10}{0.14} & \cellcolor{gray!10}{0.08} & \cellcolor{gray!10}{0.12} & \cellcolor{gray!10}{0.08} & \cellcolor{gray!10}{0.13} & \cellcolor{gray!10}{0.04} & \cellcolor{gray!10}{0.08} & \cellcolor{gray!10}{0.06}\\
\bottomrule
\end{longtable}
\endgroup{}

\newpage

\begingroup\fontsize{7}{9}\selectfont

\begin{longtable}[t]{>{\raggedright\arraybackslash}p{0.5in}>{\raggedleft\arraybackslash}p{0.325in}>{\raggedleft\arraybackslash}p{0.325in}>{\raggedleft\arraybackslash}p{0.325in}>{\raggedleft\arraybackslash}p{0.325in}>{\raggedleft\arraybackslash}p{0.325in}>{\raggedleft\arraybackslash}p{0.325in}>{\raggedleft\arraybackslash}p{0.325in}>{\raggedleft\arraybackslash}p{0.325in}}
\caption{Pecent of 95 percent Bayesian credible intervals that covered the true
effective population size, from all inferential techniques in each data
scenario for 500 simulations in scenarios A, B, and C. Time periods span
approximately from first simulated sampling time to 90th percentil of
historic reporting delays, 41 days. The final time period for each
simulation scenario spans from the first to last sampling time
simulated.}\tabularnewline

\toprule
\multicolumn{1}{c}{ } & \multicolumn{2}{c}{Retrospective} & \multicolumn{6}{c}{Real-time inference} \\
\cmidrule(l{3pt}r{3pt}){2-3} \cmidrule(l{3pt}r{3pt}){4-9}
Time period (days) & BNPR & BNPR PS & Trunc. BNPR & Tunc. BNPR PS & BNPR & BNPR PS & BNPR PS with RP offset & BNPR PS with RP covariate\\
\midrule
\addlinespace[0.3em]
\multicolumn{9}{l}{\textbf{Scenario A}}\\
\hspace{1em}\cellcolor{gray!10}{{}[0,7)} & \cellcolor{gray!10}{98.76} & \cellcolor{gray!10}{98.04} & \cellcolor{gray!10}{} & \cellcolor{gray!10}{} & \cellcolor{gray!10}{100.00} & \cellcolor{gray!10}{0.00} & \cellcolor{gray!10}{100.00} & \cellcolor{gray!10}{66.67}\\
\hspace{1em}{}[7,14) & 99.58 & 98.46 &  &  & 96.13 & 1.62 & 96.59 & 78.69\\
\hspace{1em}\cellcolor{gray!10}{{}[14,21)} & \cellcolor{gray!10}{99.61} & \cellcolor{gray!10}{98.86} & \cellcolor{gray!10}{} & \cellcolor{gray!10}{} & \cellcolor{gray!10}{98.44} & \cellcolor{gray!10}{69.31} & \cellcolor{gray!10}{97.42} & \cellcolor{gray!10}{94.37}\\
\hspace{1em}{}[21,28) & 99.71 & 98.79 &  &  & 99.52 & 98.14 & 98.10 & 98.52\\
\hspace{1em}\cellcolor{gray!10}{{}[28,35)} & \cellcolor{gray!10}{99.38} & \cellcolor{gray!10}{98.98} & \cellcolor{gray!10}{} & \cellcolor{gray!10}{} & \cellcolor{gray!10}{99.28} & \cellcolor{gray!10}{96.37} & \cellcolor{gray!10}{98.69} & \cellcolor{gray!10}{98.55}\\
\hspace{1em}{}[35,42) & 99.10 & 98.86 & 67.90 & 91.65 & 99.03 & 95.64 & 98.58 & 98.57\\
\hspace{1em}\cellcolor{gray!10}{{}[0,154]} & \cellcolor{gray!10}{98.75} & \cellcolor{gray!10}{96.79} & \cellcolor{gray!10}{97.21} & \cellcolor{gray!10}{95.84} & \cellcolor{gray!10}{98.71} & \cellcolor{gray!10}{90.96} & \cellcolor{gray!10}{96.71} & \cellcolor{gray!10}{96.32}\\
\addlinespace[0.3em]
\multicolumn{9}{l}{\textbf{Scenario B}}\\
\hspace{1em}{}[7,14) & 99.13 & 99.53 &  &  & 100.00 & 3.98 & 98.62 & 99.95\\
\hspace{1em}\cellcolor{gray!10}{{}[14,21)} & \cellcolor{gray!10}{99.40} & \cellcolor{gray!10}{98.99} & \cellcolor{gray!10}{} & \cellcolor{gray!10}{} & \cellcolor{gray!10}{98.75} & \cellcolor{gray!10}{23.54} & \cellcolor{gray!10}{99.07} & \cellcolor{gray!10}{99.24}\\
\hspace{1em}{}[21,28) & 99.47 & 99.08 &  &  & 99.19 & 76.29 & 99.24 & 98.40\\
\hspace{1em}\cellcolor{gray!10}{{}[28,35)} & \cellcolor{gray!10}{99.60} & \cellcolor{gray!10}{98.97} & \cellcolor{gray!10}{} & \cellcolor{gray!10}{} & \cellcolor{gray!10}{99.62} & \cellcolor{gray!10}{95.20} & \cellcolor{gray!10}{99.25} & \cellcolor{gray!10}{99.08}\\
\hspace{1em}{}[35,42) & 99.33 & 99.03 & 79.24 & 98.40 & 99.45 & 98.39 & 99.19 & 99.18\\
\hspace{1em}\cellcolor{gray!10}{{}[0,228]} & \cellcolor{gray!10}{98.06} & \cellcolor{gray!10}{96.21} & \cellcolor{gray!10}{97.14} & \cellcolor{gray!10}{96.20} & \cellcolor{gray!10}{98.21} & \cellcolor{gray!10}{89.79} & \cellcolor{gray!10}{96.53} & \cellcolor{gray!10}{96.49}\\
\addlinespace[0.3em]
\multicolumn{9}{l}{\textbf{Scenario C}}\\
\hspace{1em}{}[7,14) & 99.35 & 98.52 &  &  & 99.24 & 0.00 & 98.76 & 96.10\\
\hspace{1em}\cellcolor{gray!10}{{}[14,21)} & \cellcolor{gray!10}{99.19} & \cellcolor{gray!10}{98.22} & \cellcolor{gray!10}{} & \cellcolor{gray!10}{} & \cellcolor{gray!10}{96.71} & \cellcolor{gray!10}{22.96} & \cellcolor{gray!10}{98.34} & \cellcolor{gray!10}{93.92}\\
\hspace{1em}{}[21,28) & 99.26 & 98.66 &  &  & 98.83 & 85.93 & 98.30 & 97.53\\
\hspace{1em}\cellcolor{gray!10}{{}[28,35)} & \cellcolor{gray!10}{99.67} & \cellcolor{gray!10}{98.79} & \cellcolor{gray!10}{} & \cellcolor{gray!10}{} & \cellcolor{gray!10}{99.40} & \cellcolor{gray!10}{98.11} & \cellcolor{gray!10}{98.51} & \cellcolor{gray!10}{98.56}\\
\hspace{1em}{}[35,42) & 99.57 & 98.50 & 23.85 & 87.97 & 99.51 & 98.36 & 98.23 & 98.34\\
\hspace{1em}\cellcolor{gray!10}{{}[0,307]} & \cellcolor{gray!10}{97.60} & \cellcolor{gray!10}{95.89} & \cellcolor{gray!10}{95.36} & \cellcolor{gray!10}{94.39} & \cellcolor{gray!10}{97.95} & \cellcolor{gray!10}{92.67} & \cellcolor{gray!10}{96.00} & \cellcolor{gray!10}{96.30}\\
\bottomrule
\end{longtable}
\endgroup{}

\newpage

\begingroup\fontsize{7}{9}\selectfont

\begin{longtable}[t]{>{\raggedright\arraybackslash}p{0.5in}>{\raggedleft\arraybackslash}p{0.325in}>{\raggedleft\arraybackslash}p{0.325in}>{\raggedleft\arraybackslash}p{0.325in}>{\raggedleft\arraybackslash}p{0.325in}>{\raggedleft\arraybackslash}p{0.325in}>{\raggedleft\arraybackslash}p{0.325in}>{\raggedleft\arraybackslash}p{0.325in}>{\raggedleft\arraybackslash}p{0.325in}}
\caption{Mean width of 95 percent Bayesian credible intervals of effective
population size from all inferential techniques in each data scenario
for 500 simulations in scenarios A, B, and C. Time periods span
approximately from first simulated sampling time to 90th percentil of
historic reporting delays, 41 days. The final time period for each
simulation scenario spans from the first to last sampling time
simulated.}\tabularnewline

\toprule
\multicolumn{1}{c}{ } & \multicolumn{2}{c}{Retrospective} & \multicolumn{6}{c}{Real-time inference} \\
\cmidrule(l{3pt}r{3pt}){2-3} \cmidrule(l{3pt}r{3pt}){4-9}
Time period (days) & BNPR & BNPR PS & Trunc. BNPR & Tunc. BNPR PS & BNPR & BNPR PS & BNPR PS with RP offset & BNPR PS with RP covariate\\
\midrule
\addlinespace[0.3em]
\multicolumn{9}{l}{\textbf{Scenario A}}\\
\hspace{1em}\cellcolor{gray!10}{{}[0,7)} & \cellcolor{gray!10}{39.06} & \cellcolor{gray!10}{23.68} & \cellcolor{gray!10}{} & \cellcolor{gray!10}{} & \cellcolor{gray!10}{138.99} & \cellcolor{gray!10}{16.69} & \cellcolor{gray!10}{83.81} & \cellcolor{gray!10}{74.24}\\
\hspace{1em}{}[7,14) & 29.61 & 19.29 &  &  & 69.37 & 21.83 & 38.36 & 44.73\\
\hspace{1em}\cellcolor{gray!10}{{}[14,21)} & \cellcolor{gray!10}{27.18} & \cellcolor{gray!10}{17.09} & \cellcolor{gray!10}{} & \cellcolor{gray!10}{} & \cellcolor{gray!10}{36.48} & \cellcolor{gray!10}{23.90} & \cellcolor{gray!10}{25.08} & \cellcolor{gray!10}{25.85}\\
\hspace{1em}{}[21,28) & 24.75 & 15.41 &  &  & 29.31 & 22.69 & 19.60 & 20.60\\
\hspace{1em}\cellcolor{gray!10}{{}[28,35)} & \cellcolor{gray!10}{22.63} & \cellcolor{gray!10}{14.10} & \cellcolor{gray!10}{} & \cellcolor{gray!10}{} & \cellcolor{gray!10}{25.77} & \cellcolor{gray!10}{20.27} & \cellcolor{gray!10}{16.60} & \cellcolor{gray!10}{17.88}\\
\hspace{1em}{}[35,42) & 20.67 & 12.95 & 31.91 & 23.54 & 23.06 & 17.94 & 14.50 & 15.69\\
\hspace{1em}\cellcolor{gray!10}{{}[0,154]} & \cellcolor{gray!10}{17.74} & \cellcolor{gray!10}{10.73} & \cellcolor{gray!10}{16.35} & \cellcolor{gray!10}{10.00} & \cellcolor{gray!10}{20.21} & \cellcolor{gray!10}{12.57} & \cellcolor{gray!10}{12.37} & \cellcolor{gray!10}{13.00}\\
\addlinespace[0.3em]
\multicolumn{9}{l}{\textbf{Scenario B}}\\
\hspace{1em}{}[7,14) & 34.39 & 19.47 &  &  & 86.69 & 20.78 & 33.97 & 51.30\\
\hspace{1em}\cellcolor{gray!10}{{}[14,21)} & \cellcolor{gray!10}{33.87} & \cellcolor{gray!10}{19.78} & \cellcolor{gray!10}{} & \cellcolor{gray!10}{} & \cellcolor{gray!10}{61.89} & \cellcolor{gray!10}{19.65} & \cellcolor{gray!10}{25.46} & \cellcolor{gray!10}{36.15}\\
\hspace{1em}{}[21,28) & 35.36 & 20.72 &  &  & 45.16 & 21.07 & 23.40 & 26.70\\
\hspace{1em}\cellcolor{gray!10}{{}[28,35)} & \cellcolor{gray!10}{37.55} & \cellcolor{gray!10}{22.12} & \cellcolor{gray!10}{} & \cellcolor{gray!10}{} & \cellcolor{gray!10}{41.90} & \cellcolor{gray!10}{22.89} & \cellcolor{gray!10}{23.98} & \cellcolor{gray!10}{24.98}\\
\hspace{1em}{}[35,42) & 39.98 & 23.83 & 64.73 & 33.94 & 42.89 & 24.92 & 25.28 & 25.56\\
\hspace{1em}\cellcolor{gray!10}{{}[0,228]} & \cellcolor{gray!10}{30.83} & \cellcolor{gray!10}{18.40} & \cellcolor{gray!10}{31.57} & \cellcolor{gray!10}{18.66} & \cellcolor{gray!10}{33.77} & \cellcolor{gray!10}{19.48} & \cellcolor{gray!10}{19.54} & \cellcolor{gray!10}{20.67}\\
\addlinespace[0.3em]
\multicolumn{9}{l}{\textbf{Scenario C}}\\
\hspace{1em}{}[7,14) & 71.12 & 45.30 &  &  & 174.69 & 42.27 & 79.11 & 114.94\\
\hspace{1em}\cellcolor{gray!10}{{}[14,21)} & \cellcolor{gray!10}{69.30} & \cellcolor{gray!10}{46.56} & \cellcolor{gray!10}{} & \cellcolor{gray!10}{} & \cellcolor{gray!10}{107.61} & \cellcolor{gray!10}{46.78} & \cellcolor{gray!10}{60.30} & \cellcolor{gray!10}{73.87}\\
\hspace{1em}{}[21,28) & 67.87 & 45.53 &  &  & 79.86 & 51.78 & 53.39 & 55.21\\
\hspace{1em}\cellcolor{gray!10}{{}[28,35)} & \cellcolor{gray!10}{64.83} & \cellcolor{gray!10}{43.03} & \cellcolor{gray!10}{} & \cellcolor{gray!10}{} & \cellcolor{gray!10}{71.50} & \cellcolor{gray!10}{50.95} & \cellcolor{gray!10}{48.45} & \cellcolor{gray!10}{49.27}\\
\hspace{1em}{}[35,42) & 61.57 & 39.92 & 73.03 & 53.62 & 65.93 & 47.69 & 43.52 & 44.63\\
\hspace{1em}\cellcolor{gray!10}{{}[0,307]} & \cellcolor{gray!10}{47.19} & \cellcolor{gray!10}{28.39} & \cellcolor{gray!10}{45.18} & \cellcolor{gray!10}{27.16} & \cellcolor{gray!10}{50.51} & \cellcolor{gray!10}{31.19} & \cellcolor{gray!10}{30.12} & \cellcolor{gray!10}{31.94}\\
\bottomrule
\end{longtable}
\endgroup{}

\newpage

\begin{figure}[H]

{\centering \includegraphics[width=\textwidth]{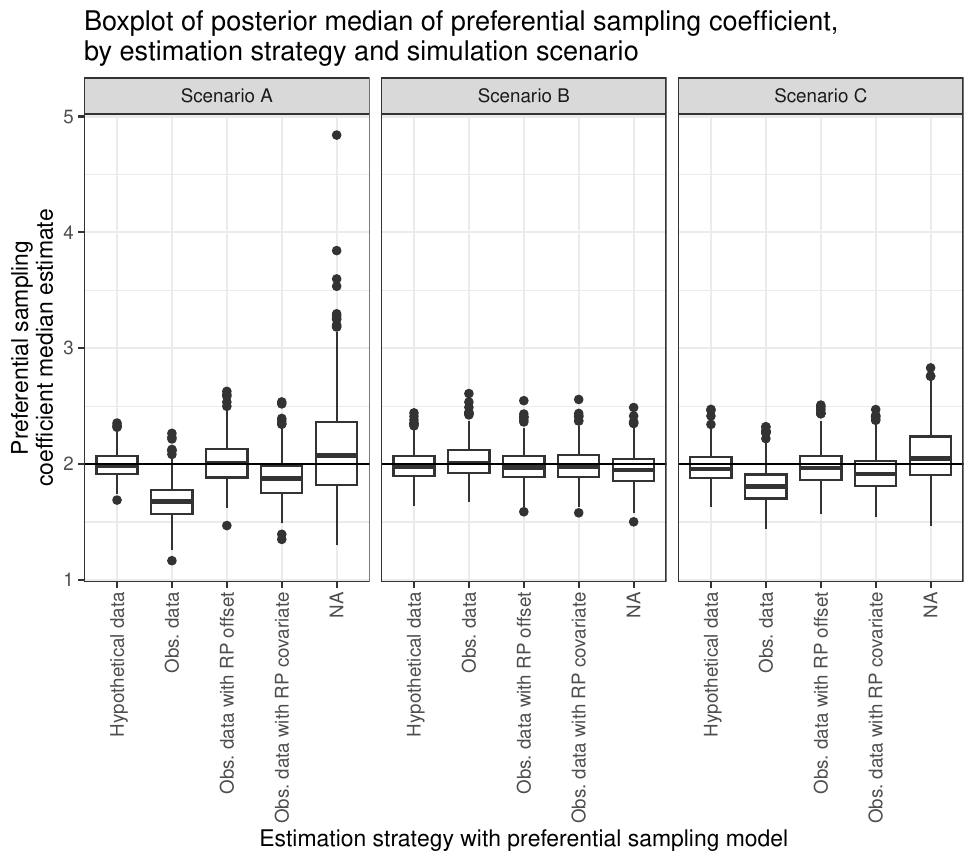}

}

\caption{Median preferential sampling coeffecient estimate from
simulations by estimation strategy and simulation scenario.}

\end{figure}%

\newpage{}

\subsection{Simulation with Santa Clara County reporting delays (more
extreme delays than Washington
state)}\label{simulation-with-santa-clara-county-reporting-delays-more-extreme-delays-than-washington-state}

All simulation details are identical to the simulations in the main
manuscript, with a difference in reporting probabilities. Here we use
the empirical reporting delays of Santa Clara County as the reporting
probabilities. The 90th percentile of reporting delays is 54 for Santa
Clara County, while Washington state's was 41. Since these delays are
more extreme this serves to help investigate how more extreme delays
affect the analysis.

\subsubsection{Results from last simulation in each
scenario}\label{results-from-last-simulation-in-each-scenario-1}

\begin{figure}[H]

\centering{

\includegraphics[width=\textwidth]{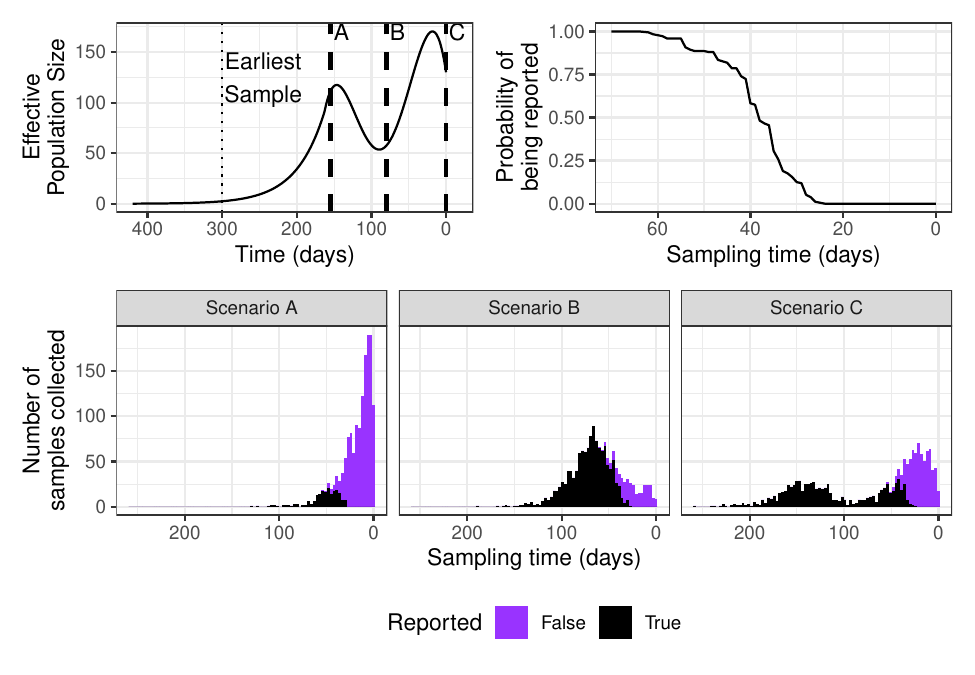}

}

\caption{\label{fig-simulation-set-up-plot}Three panel plot providing
simulation details: effective population trajectories (upper left plot),
reporting probability by sampling time (upper right plot) obtained from
Santa Clara County empirial cumulative distibution, and histograms of
sampling times from the last simulation of in each simulation scenario
colored by whether sample was reported by time of analysis (bottom
plots). Each simulation scenario had a different time zero, i.e.~time of
latest sample (dashed lines). The earliest sampling time in each
scenario was at the same point in the trajectory (dotted line).}

\end{figure}%

\begin{figure}[H]

{\centering \includegraphics[width=\textwidth]{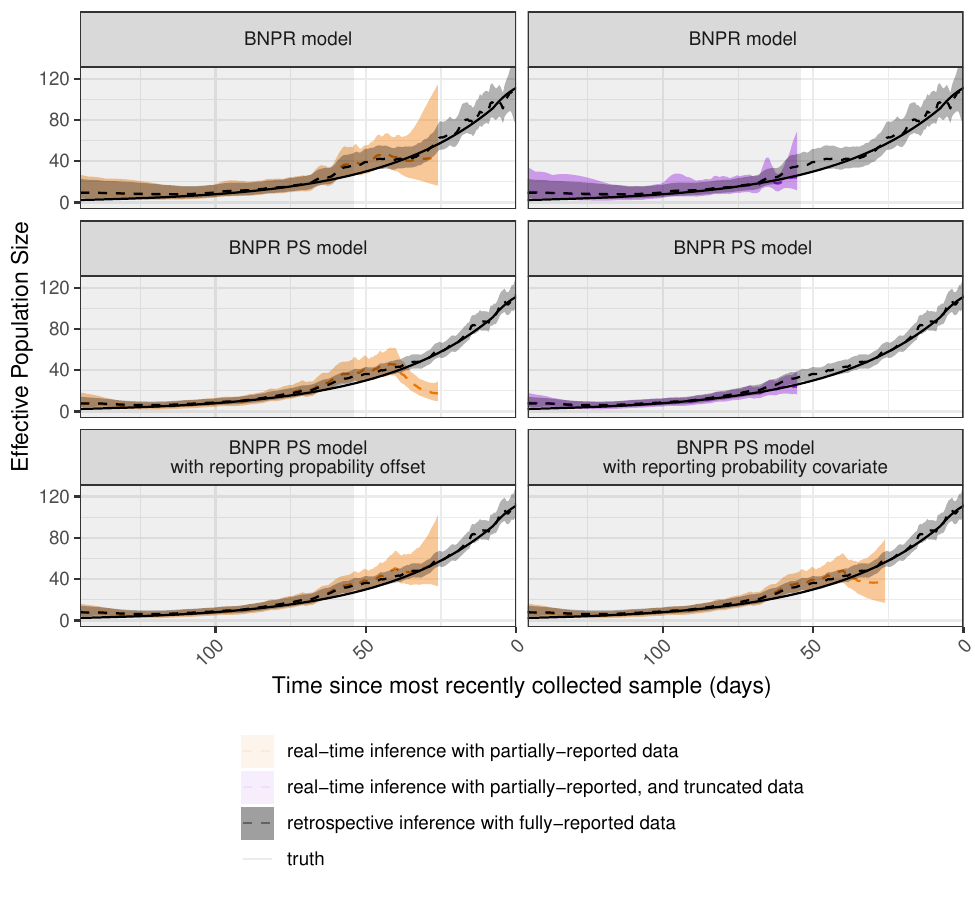}

}

\caption{Comparison of phylodynamic estimation methods of effective
population size trajectory for three different simulated data scenarios
from the scenario A trajectory.}

\end{figure}%

\begin{figure}[H]

{\centering \includegraphics[width=\textwidth]{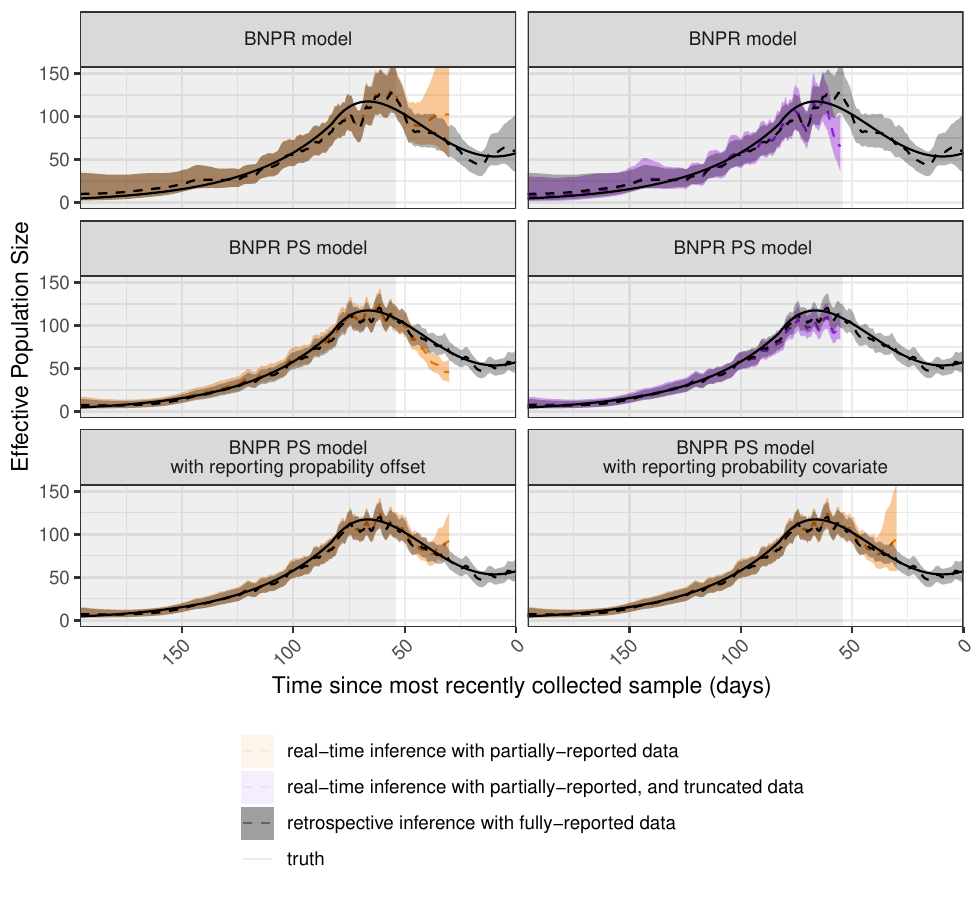}

}

\caption{Comparison of phylodynamic estimation methods of effective
population size trajectory for three different simulated data scenarios
from the scenario B trajectory.}

\end{figure}%

\begin{figure}[H]

{\centering \includegraphics[width=\textwidth]{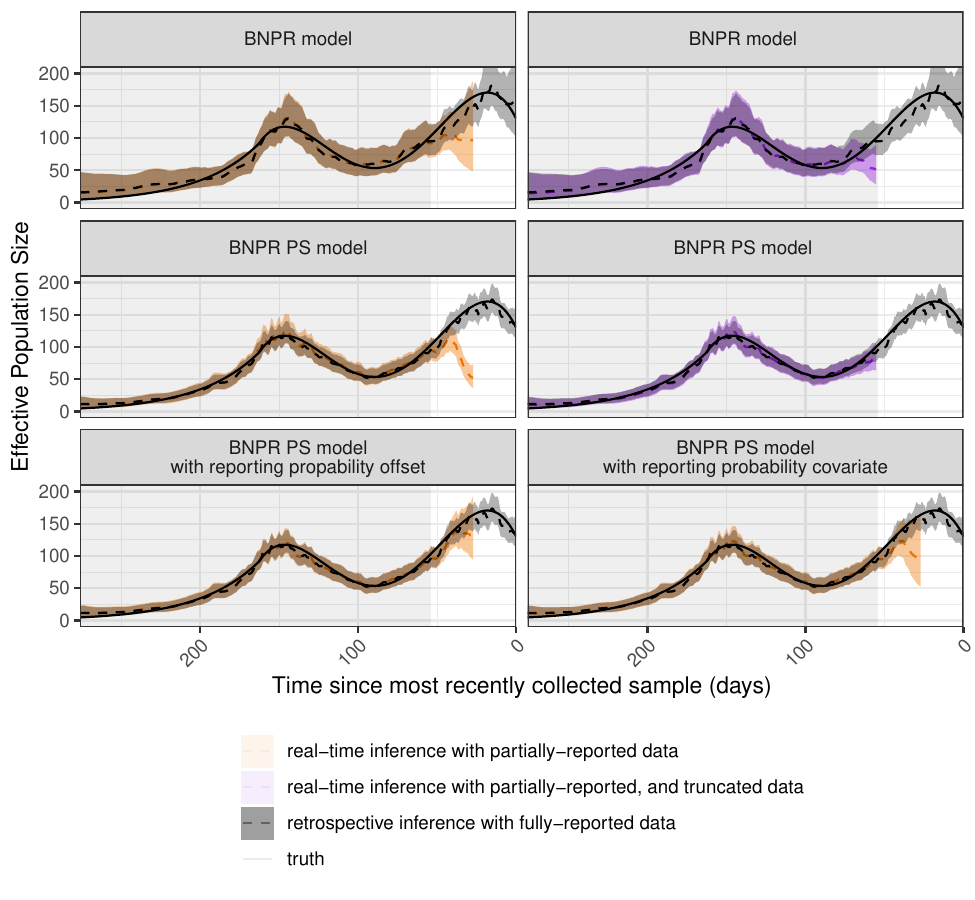}

}

\caption{Comparison of phylodynamic estimation methods of effective
population size trajectory for three different simulated data scenarios
from the scenario C trajectory.}

\end{figure}%

\newpage

\subsubsection{Performance metrics across all
simulations}\label{performance-metrics-across-all-simulations-1}

\begingroup\fontsize{7}{9}\selectfont

\begin{longtable}[t]{>{\raggedright\arraybackslash}p{0.5in}>{\raggedleft\arraybackslash}p{0.325in}>{\raggedleft\arraybackslash}p{0.325in}>{\raggedleft\arraybackslash}p{0.325in}>{\raggedleft\arraybackslash}p{0.325in}>{\raggedleft\arraybackslash}p{0.325in}>{\raggedleft\arraybackslash}p{0.325in}>{\raggedleft\arraybackslash}p{0.325in}>{\raggedleft\arraybackslash}p{0.325in}}

\caption{\label{tbl-mrd}Mean relative deviation of effective population
size from all estimation techniques in each data scenario for 500
simulations in scenarios A, B, and C. Time periods span from first
simulated sampling time to 90th percentil of historic reporting delays,
55 days. The final time period for each simulation scenario spans from
the first to last sampling time simulated.}

\tabularnewline

\toprule
\multicolumn{1}{c}{ } & \multicolumn{2}{c}{Retrospective} & \multicolumn{6}{c}{Real-time inference} \\
\cmidrule(l{3pt}r{3pt}){2-3} \cmidrule(l{3pt}r{3pt}){4-9}
Time period (days) & BNPR & BNPR PS & Trunc. BNPR & Tunc. BNPR PS & BNPR & BNPR PS & BNPR PS with RP offset & BNPR PS with RP covariate\\
\midrule
\addlinespace[0.3em]
\multicolumn{9}{l}{\textbf{Scenario A}}\\
\hspace{1em}\cellcolor{gray!10}{{}[0,7)} & \cellcolor{gray!10}{-0.01} & \cellcolor{gray!10}{0.00} & \cellcolor{gray!10}{} & \cellcolor{gray!10}{} & \cellcolor{gray!10}{-0.36} & \cellcolor{gray!10}{-0.96} & \cellcolor{gray!10}{0.07} & \cellcolor{gray!10}{-0.50}\\
\hspace{1em}{}[7,14) & 0.01 & 0.00 &  &  & -0.46 & -0.95 & 0.00 & -0.53\\
\hspace{1em}\cellcolor{gray!10}{{}[14,21)} & \cellcolor{gray!10}{0.00} & \cellcolor{gray!10}{0.00} & \cellcolor{gray!10}{} & \cellcolor{gray!10}{} & \cellcolor{gray!10}{-0.38} & \cellcolor{gray!10}{-0.90} & \cellcolor{gray!10}{-0.02} & \cellcolor{gray!10}{-0.50}\\
\hspace{1em}{}[21,28) & 0.00 & 0.00 &  &  & -0.23 & -0.68 & -0.04 & -0.35\\
\hspace{1em}\cellcolor{gray!10}{{}[28,35)} & \cellcolor{gray!10}{0.00} & \cellcolor{gray!10}{-0.01} & \cellcolor{gray!10}{} & \cellcolor{gray!10}{} & \cellcolor{gray!10}{-0.14} & \cellcolor{gray!10}{-0.51} & \cellcolor{gray!10}{-0.02} & \cellcolor{gray!10}{-0.24}\\
\hspace{1em}{}[35,42) & 0.00 & -0.01 &  &  & -0.03 & -0.16 & 0.00 & -0.06\\
\hspace{1em}\cellcolor{gray!10}{{}[42,49)} & \cellcolor{gray!10}{0.01} & \cellcolor{gray!10}{-0.01} & \cellcolor{gray!10}{} & \cellcolor{gray!10}{} & \cellcolor{gray!10}{0.01} & \cellcolor{gray!10}{0.04} & \cellcolor{gray!10}{0.01} & \cellcolor{gray!10}{0.02}\\
\hspace{1em}{}[49,56) & 0.01 & -0.01 & -0.35 & -0.12 & 0.01 & 0.06 & -0.01 & 0.02\\
\hspace{1em}\cellcolor{gray!10}{{}[0,156]} & \cellcolor{gray!10}{0.12} & \cellcolor{gray!10}{0.05} & \cellcolor{gray!10}{0.11} & \cellcolor{gray!10}{0.02} & \cellcolor{gray!10}{0.10} & \cellcolor{gray!10}{-0.01} & \cellcolor{gray!10}{0.04} & \cellcolor{gray!10}{0.01}\\
\addlinespace[0.3em]
\multicolumn{9}{l}{\textbf{Scenario B}}\\
\hspace{1em}{}[21,28) & 0.01 & 0.00 &  &  & 0.39 & -0.54 & 0.20 & 0.18\\
\hspace{1em}\cellcolor{gray!10}{{}[28,35)} & \cellcolor{gray!10}{0.00} & \cellcolor{gray!10}{0.00} & \cellcolor{gray!10}{} & \cellcolor{gray!10}{} & \cellcolor{gray!10}{0.22} & \cellcolor{gray!10}{-0.48} & \cellcolor{gray!10}{0.06} & \cellcolor{gray!10}{0.07}\\
\hspace{1em}{}[35,42) & 0.00 & 0.00 &  &  & 0.08 & -0.27 & -0.01 & 0.00\\
\hspace{1em}\cellcolor{gray!10}{{}[42,49)} & \cellcolor{gray!10}{0.00} & \cellcolor{gray!10}{0.00} & \cellcolor{gray!10}{} & \cellcolor{gray!10}{} & \cellcolor{gray!10}{0.01} & \cellcolor{gray!10}{-0.08} & \cellcolor{gray!10}{0.00} & \cellcolor{gray!10}{0.00}\\
\hspace{1em}{}[49,56) & 0.00 & 0.00 & -0.36 & -0.07 & 0.00 & -0.03 & 0.00 & 0.00\\
\hspace{1em}\cellcolor{gray!10}{{}[0,229]} & \cellcolor{gray!10}{0.15} & \cellcolor{gray!10}{0.06} & \cellcolor{gray!10}{0.13} & \cellcolor{gray!10}{0.08} & \cellcolor{gray!10}{0.17} & \cellcolor{gray!10}{0.06} & \cellcolor{gray!10}{0.07} & \cellcolor{gray!10}{0.07}\\
\addlinespace[0.3em]
\multicolumn{9}{l}{\textbf{Scenario C}}\\
\hspace{1em}{}[14,21) & -0.01 & 0.01 &  &  & -0.06 & -0.84 & 0.18 & -0.26\\
\hspace{1em}\cellcolor{gray!10}{{}[21,28)} & \cellcolor{gray!10}{0.00} & \cellcolor{gray!10}{0.01} & \cellcolor{gray!10}{} & \cellcolor{gray!10}{} & \cellcolor{gray!10}{-0.18} & \cellcolor{gray!10}{-0.67} & \cellcolor{gray!10}{-0.03} & \cellcolor{gray!10}{-0.27}\\
\hspace{1em}{}[28,35) & 0.00 & 0.00 &  &  & -0.16 & -0.55 & -0.04 & -0.23\\
\hspace{1em}\cellcolor{gray!10}{{}[35,42)} & \cellcolor{gray!10}{0.00} & \cellcolor{gray!10}{0.00} & \cellcolor{gray!10}{} & \cellcolor{gray!10}{} & \cellcolor{gray!10}{-0.08} & \cellcolor{gray!10}{-0.28} & \cellcolor{gray!10}{-0.01} & \cellcolor{gray!10}{-0.10}\\
\hspace{1em}{}[42,49) & 0.00 & 0.00 &  &  & -0.03 & -0.08 & 0.00 & -0.02\\
\hspace{1em}\cellcolor{gray!10}{{}[49,56)} & \cellcolor{gray!10}{0.00} & \cellcolor{gray!10}{0.00} & \cellcolor{gray!10}{-0.46} & \cellcolor{gray!10}{-0.11} & \cellcolor{gray!10}{0.00} & \cellcolor{gray!10}{-0.02} & \cellcolor{gray!10}{0.00} & \cellcolor{gray!10}{-0.01}\\
\hspace{1em}{}[0,307] & 0.14 & 0.07 & 0.12 & 0.05 & 0.13 & 0.04 & 0.07 & 0.05\\
\bottomrule

\end{longtable}

\endgroup{}

\newpage

\begingroup\fontsize{7}{9}\selectfont

\begin{longtable}[t]{>{\raggedright\arraybackslash}p{0.5in}>{\raggedleft\arraybackslash}p{0.325in}>{\raggedleft\arraybackslash}p{0.325in}>{\raggedleft\arraybackslash}p{0.325in}>{\raggedleft\arraybackslash}p{0.325in}>{\raggedleft\arraybackslash}p{0.325in}>{\raggedleft\arraybackslash}p{0.325in}>{\raggedleft\arraybackslash}p{0.325in}>{\raggedleft\arraybackslash}p{0.325in}}

\caption{\label{tbl-rpc}Pecent of 95 percent Bayesian credible intervals
that covered the true effective population size, from all estimation
techniques in each data scenario for 500 simulations in scenarios A, B,
and C. Time periods span from first simulated sampling time to 90th
percentil of historic reporting delays, 55 days. The final time period
for each simulation scenario spans from the first to last sampling time
simulated.}

\tabularnewline

\toprule
\multicolumn{1}{c}{ } & \multicolumn{2}{c}{Retrospective} & \multicolumn{6}{c}{Real-time inference} \\
\cmidrule(l{3pt}r{3pt}){2-3} \cmidrule(l{3pt}r{3pt}){4-9}
Time period (days) & BNPR & BNPR PS & Trunc. BNPR & Tunc. BNPR PS & BNPR & BNPR PS & BNPR PS with RP offset & BNPR PS with RP covariate\\
\midrule
\addlinespace[0.3em]
\multicolumn{9}{l}{\textbf{Scenario A}}\\
\hspace{1em}\cellcolor{gray!10}{{}[0,7)} & \cellcolor{gray!10}{99.13} & \cellcolor{gray!10}{97.97} & \cellcolor{gray!10}{} & \cellcolor{gray!10}{} & \cellcolor{gray!10}{100.00} & \cellcolor{gray!10}{0.00} & \cellcolor{gray!10}{100.00} & \cellcolor{gray!10}{100.00}\\
\hspace{1em}{}[7,14) & 99.20 & 98.48 &  &  & 100.00 & 0.00 & 100.00 & 100.00\\
\hspace{1em}\cellcolor{gray!10}{{}[14,21)} & \cellcolor{gray!10}{99.25} & \cellcolor{gray!10}{98.75} & \cellcolor{gray!10}{} & \cellcolor{gray!10}{} & \cellcolor{gray!10}{100.00} & \cellcolor{gray!10}{0.00} & \cellcolor{gray!10}{100.00} & \cellcolor{gray!10}{90.91}\\
\hspace{1em}{}[21,28) & 99.19 & 98.74 &  &  & 99.20 & 0.00 & 99.70 & 90.16\\
\hspace{1em}\cellcolor{gray!10}{{}[28,35)} & \cellcolor{gray!10}{99.45} & \cellcolor{gray!10}{99.12} & \cellcolor{gray!10}{} & \cellcolor{gray!10}{} & \cellcolor{gray!10}{97.80} & \cellcolor{gray!10}{3.30} & \cellcolor{gray!10}{98.05} & \cellcolor{gray!10}{88.91}\\
\hspace{1em}{}[35,42) & 99.54 & 98.88 &  &  & 98.56 & 74.88 & 98.26 & 95.61\\
\hspace{1em}\cellcolor{gray!10}{{}[42,49)} & \cellcolor{gray!10}{99.46} & \cellcolor{gray!10}{98.55} & \cellcolor{gray!10}{} & \cellcolor{gray!10}{} & \cellcolor{gray!10}{99.57} & \cellcolor{gray!10}{98.32} & \cellcolor{gray!10}{98.11} & \cellcolor{gray!10}{98.51}\\
\hspace{1em}{}[49,56) & 99.13 & 98.56 & 81.81 & 93.91 & 99.23 & 98.15 & 98.47 & 98.69\\
\hspace{1em}\cellcolor{gray!10}{{}[0,156]} & \cellcolor{gray!10}{98.36} & \cellcolor{gray!10}{96.71} & \cellcolor{gray!10}{97.42} & \cellcolor{gray!10}{96.68} & \cellcolor{gray!10}{98.83} & \cellcolor{gray!10}{89.72} & \cellcolor{gray!10}{97.57} & \cellcolor{gray!10}{97.10}\\
\addlinespace[0.3em]
\multicolumn{9}{l}{\textbf{Scenario B}}\\
\hspace{1em}{}[21,28) & 99.70 & 99.62 &  &  & 100.00 & 0.00 & 96.04 & 100.00\\
\hspace{1em}\cellcolor{gray!10}{{}[28,35)} & \cellcolor{gray!10}{99.45} & \cellcolor{gray!10}{99.35} & \cellcolor{gray!10}{} & \cellcolor{gray!10}{} & \cellcolor{gray!10}{99.95} & \cellcolor{gray!10}{0.00} & \cellcolor{gray!10}{99.02} & \cellcolor{gray!10}{99.98}\\
\hspace{1em}{}[35,42) & 99.27 & 98.84 &  &  & 98.89 & 14.36 & 98.89 & 98.90\\
\hspace{1em}\cellcolor{gray!10}{{}[42,49)} & \cellcolor{gray!10}{99.09} & \cellcolor{gray!10}{99.02} & \cellcolor{gray!10}{} & \cellcolor{gray!10}{} & \cellcolor{gray!10}{99.11} & \cellcolor{gray!10}{85.62} & \cellcolor{gray!10}{98.99} & \cellcolor{gray!10}{98.64}\\
\hspace{1em}{}[49,56) & 99.25 & 99.14 & 44.05 & 92.03 & 99.23 & 97.53 & 99.04 & 99.11\\
\hspace{1em}\cellcolor{gray!10}{{}[0,229]} & \cellcolor{gray!10}{97.83} & \cellcolor{gray!10}{96.38} & \cellcolor{gray!10}{96.09} & \cellcolor{gray!10}{94.74} & \cellcolor{gray!10}{98.05} & \cellcolor{gray!10}{88.56} & \cellcolor{gray!10}{96.44} & \cellcolor{gray!10}{96.49}\\
\addlinespace[0.3em]
\multicolumn{9}{l}{\textbf{Scenario C}}\\
\hspace{1em}{}[14,21) & 98.72 & 98.54 &  &  & 100.00 & 0.00 & 100.00 & 100.00\\
\hspace{1em}\cellcolor{gray!10}{{}[21,28)} & \cellcolor{gray!10}{99.80} & \cellcolor{gray!10}{99.21} & \cellcolor{gray!10}{} & \cellcolor{gray!10}{} & \cellcolor{gray!10}{99.53} & \cellcolor{gray!10}{0.00} & \cellcolor{gray!10}{99.37} & \cellcolor{gray!10}{95.91}\\
\hspace{1em}{}[28,35) & 99.31 & 98.59 &  &  & 98.27 & 0.00 & 98.33 & 89.99\\
\hspace{1em}\cellcolor{gray!10}{{}[35,42)} & \cellcolor{gray!10}{99.50} & \cellcolor{gray!10}{98.87} & \cellcolor{gray!10}{} & \cellcolor{gray!10}{} & \cellcolor{gray!10}{97.12} & \cellcolor{gray!10}{20.17} & \cellcolor{gray!10}{98.43} & \cellcolor{gray!10}{91.60}\\
\hspace{1em}{}[42,49) & 99.68 & 99.50 &  &  & 99.19 & 90.77 & 98.82 & 98.35\\
\hspace{1em}\cellcolor{gray!10}{{}[49,56)} & \cellcolor{gray!10}{99.19} & \cellcolor{gray!10}{98.97} & \cellcolor{gray!10}{32.32} & \cellcolor{gray!10}{92.99} & \cellcolor{gray!10}{99.42} & \cellcolor{gray!10}{98.84} & \cellcolor{gray!10}{99.08} & \cellcolor{gray!10}{99.16}\\
\hspace{1em}{}[0,307] & 97.86 & 96.15 & 95.46 & 96.48 & 98.32 & 91.67 & 96.77 & 96.61\\
\bottomrule

\end{longtable}

\endgroup{}

\newpage

\begingroup\fontsize{7}{9}\selectfont

\begin{longtable}[t]{>{\raggedright\arraybackslash}p{0.5in}>{\raggedleft\arraybackslash}p{0.325in}>{\raggedleft\arraybackslash}p{0.325in}>{\raggedleft\arraybackslash}p{0.325in}>{\raggedleft\arraybackslash}p{0.325in}>{\raggedleft\arraybackslash}p{0.325in}>{\raggedleft\arraybackslash}p{0.325in}>{\raggedleft\arraybackslash}p{0.325in}>{\raggedleft\arraybackslash}p{0.325in}}

\caption{\label{tbl-mbw}Mean width of 95 percent Bayesian credible
interval of effective population size from all estimation techniques in
each data scenario for 500 simulations in scenarios A, B, and C. Time
periods span from first simulated sampling time to 90th percentil of
historic reporting delays, 55 days. The final time period for each
simulation scenario spans from the first to last sampling time
simulated.}

\tabularnewline

\toprule
\multicolumn{1}{c}{ } & \multicolumn{2}{c}{Retrospective} & \multicolumn{6}{c}{Real-time inference} \\
\cmidrule(l{3pt}r{3pt}){2-3} \cmidrule(l{3pt}r{3pt}){4-9}
Time period (days) & BNPR & BNPR PS & Trunc. BNPR & Tunc. BNPR PS & BNPR & BNPR PS & BNPR PS with RP offset & BNPR PS with RP covariate\\
\midrule
\addlinespace[0.3em]
\multicolumn{9}{l}{\textbf{Scenario A}}\\
\hspace{1em}\cellcolor{gray!10}{{}[0,7)} & \cellcolor{gray!10}{38.90} & \cellcolor{gray!10}{23.77} & \cellcolor{gray!10}{} & \cellcolor{gray!10}{} & \cellcolor{gray!10}{530.38} & \cellcolor{gray!10}{9.86} & \cellcolor{gray!10}{583.96} & \cellcolor{gray!10}{208.03}\\
\hspace{1em}{}[7,14) & 29.59 & 19.36 &  &  & 262.33 & 8.94 & 306.41 & 120.18\\
\hspace{1em}\cellcolor{gray!10}{{}[14,21)} & \cellcolor{gray!10}{27.00} & \cellcolor{gray!10}{17.13} & \cellcolor{gray!10}{} & \cellcolor{gray!10}{} & \cellcolor{gray!10}{164.57} & \cellcolor{gray!10}{10.44} & \cellcolor{gray!10}{155.74} & \cellcolor{gray!10}{77.54}\\
\hspace{1em}{}[21,28) & 24.62 & 15.46 &  &  & 99.80 & 16.86 & 63.91 & 54.50\\
\hspace{1em}\cellcolor{gray!10}{{}[28,35)} & \cellcolor{gray!10}{22.49} & \cellcolor{gray!10}{14.08} & \cellcolor{gray!10}{} & \cellcolor{gray!10}{} & \cellcolor{gray!10}{66.95} & \cellcolor{gray!10}{17.33} & \cellcolor{gray!10}{39.84} & \cellcolor{gray!10}{41.11}\\
\hspace{1em}{}[35,42) & 20.59 & 12.94 &  &  & 34.46 & 20.61 & 24.48 & 24.51\\
\hspace{1em}\cellcolor{gray!10}{{}[42,49)} & \cellcolor{gray!10}{18.84} & \cellcolor{gray!10}{11.93} & \cellcolor{gray!10}{} & \cellcolor{gray!10}{} & \cellcolor{gray!10}{24.46} & \cellcolor{gray!10}{20.31} & \cellcolor{gray!10}{17.88} & \cellcolor{gray!10}{18.87}\\
\hspace{1em}{}[49,56) & 17.22 & 10.96 & 30.12 & 24.04 & 21.32 & 17.39 & 14.60 & 15.76\\
\hspace{1em}\cellcolor{gray!10}{{}[0,156]} & \cellcolor{gray!10}{17.60} & \cellcolor{gray!10}{10.71} & \cellcolor{gray!10}{15.91} & \cellcolor{gray!10}{9.87} & \cellcolor{gray!10}{20.76} & \cellcolor{gray!10}{12.11} & \cellcolor{gray!10}{13.27} & \cellcolor{gray!10}{13.47}\\
\addlinespace[0.3em]
\multicolumn{9}{l}{\textbf{Scenario B}}\\
\hspace{1em}{}[21,28) & 34.80 & 20.79 &  &  & 153.01 & 23.21 & 60.86 & 91.86\\
\hspace{1em}\cellcolor{gray!10}{{}[28,35)} & \cellcolor{gray!10}{36.89} & \cellcolor{gray!10}{22.20} & \cellcolor{gray!10}{} & \cellcolor{gray!10}{} & \cellcolor{gray!10}{116.57} & \cellcolor{gray!10}{22.09} & \cellcolor{gray!10}{43.01} & \cellcolor{gray!10}{74.10}\\
\hspace{1em}{}[35,42) & 39.54 & 23.89 &  &  & 72.47 & 24.03 & 30.70 & 43.77\\
\hspace{1em}\cellcolor{gray!10}{{}[42,49)} & \cellcolor{gray!10}{41.89} & \cellcolor{gray!10}{25.81} & \cellcolor{gray!10}{} & \cellcolor{gray!10}{} & \cellcolor{gray!10}{50.06} & \cellcolor{gray!10}{27.78} & \cellcolor{gray!10}{28.68} & \cellcolor{gray!10}{30.44}\\
\hspace{1em}{}[49,56) & 43.87 & 27.80 & 65.45 & 36.93 & 47.67 & 30.36 & 29.88 & 30.25\\
\hspace{1em}\cellcolor{gray!10}{{}[0,229]} & \cellcolor{gray!10}{30.73} & \cellcolor{gray!10}{18.44} & \cellcolor{gray!10}{30.96} & \cellcolor{gray!10}{18.00} & \cellcolor{gray!10}{35.40} & \cellcolor{gray!10}{20.32} & \cellcolor{gray!10}{20.10} & \cellcolor{gray!10}{21.98}\\
\addlinespace[0.3em]
\multicolumn{9}{l}{\textbf{Scenario C}}\\
\hspace{1em}{}[14,21) & 68.83 & 46.38 &  &  & 375.50 & 28.30 & 273.68 & 200.07\\
\hspace{1em}\cellcolor{gray!10}{{}[21,28)} & \cellcolor{gray!10}{67.69} & \cellcolor{gray!10}{45.33} & \cellcolor{gray!10}{} & \cellcolor{gray!10}{} & \cellcolor{gray!10}{225.11} & \cellcolor{gray!10}{38.12} & \cellcolor{gray!10}{128.97} & \cellcolor{gray!10}{147.24}\\
\hspace{1em}{}[28,35) & 64.90 & 42.95 &  &  & 169.17 & 36.12 & 88.34 & 114.28\\
\hspace{1em}\cellcolor{gray!10}{{}[35,42)} & \cellcolor{gray!10}{61.57} & \cellcolor{gray!10}{39.90} & \cellcolor{gray!10}{} & \cellcolor{gray!10}{} & \cellcolor{gray!10}{101.09} & \cellcolor{gray!10}{41.18} & \cellcolor{gray!10}{58.65} & \cellcolor{gray!10}{66.72}\\
\hspace{1em}{}[42,49) & 57.18 & 36.53 &  &  & 68.31 & 43.16 & 45.50 & 45.75\\
\hspace{1em}\cellcolor{gray!10}{{}[49,56)} & \cellcolor{gray!10}{53.04} & \cellcolor{gray!10}{33.49} & \cellcolor{gray!10}{63.02} & \cellcolor{gray!10}{48.23} & \cellcolor{gray!10}{59.05} & \cellcolor{gray!10}{39.49} & \cellcolor{gray!10}{38.77} & \cellcolor{gray!10}{39.00}\\
\hspace{1em}{}[0,307] & 47.22 & 28.35 & 44.08 & 27.42 & 50.66 & 30.03 & 30.76 & 32.15\\
\bottomrule

\end{longtable}

\endgroup{}

\newpage

\begin{figure}[H]

{\centering \includegraphics[width=\textwidth]{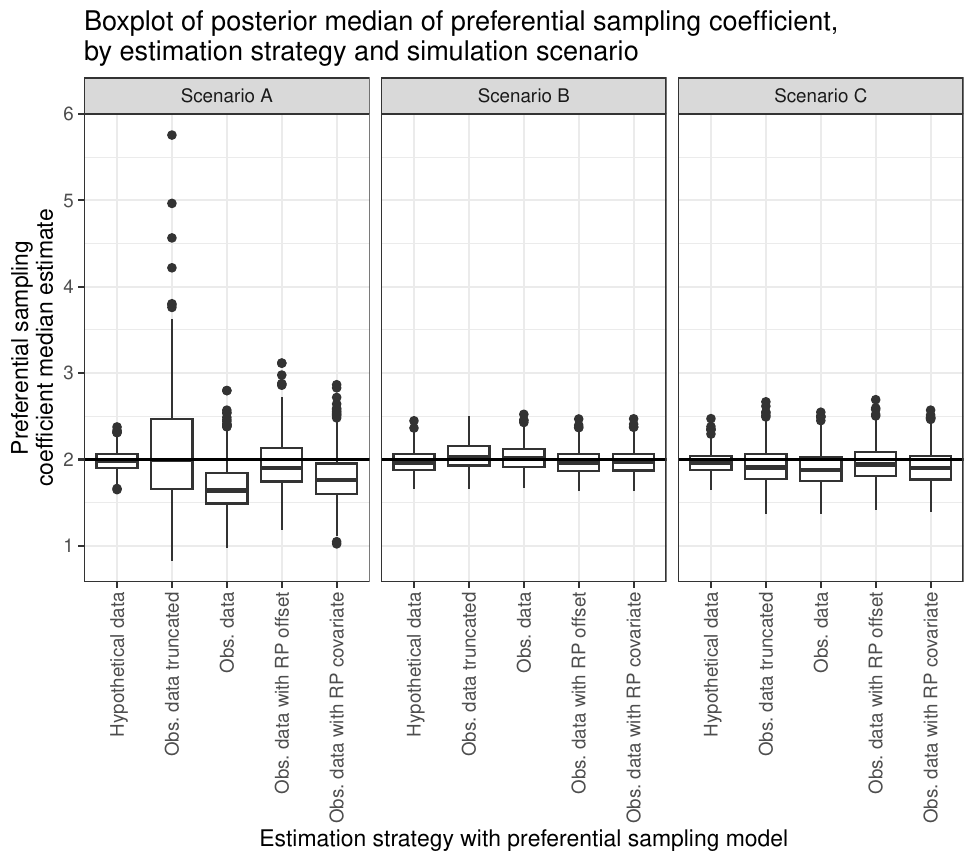}

}

\caption{Median preferential sampling coeffecient estimate from
simulations by estimation strategy and simulation scenario.}

\end{figure}%

\newpage

\section{Real data investigation: Washinton state COVID
dynamics}\label{real-data-investigation-washinton-state-covid-dynamics}

\subsection{Sequences from GISAID}\label{sequences-from-gisaid}

\textbf{Data Availability}

GISAID Identifier: EPI\_SET\_240619zd\\
doi:
\href{https://doi.org/10.55876/gis8.240619zd}{10.55876/gis8.240619zd}

All genome sequences and associated metadata in this dataset are
published in GISAID's EpiCoV database. To view the contributors of each
individual sequence with details such as accession number, Virus name,
Collection date, Originating Lab and Submitting Lab and the list of
Authors, visit
\href{https://doi.org/10.55876/gis8.240619zd}{10.55876/gis8.240619zd}

\textbf{Data Snapshot}

\begin{itemize}
\tightlist
\item
  EPI\_SET\_240619zd is composed of 500 individual genome sequences.
\item
  The collection dates range from 2021-02-01 to 2021-08-01;
\item
  Data were collected in 1 countries and territories;
\item
  All sequences in this dataset are compared relative to
  hCoV-19/Wuhan/WIV04/2019 (WIV04), the official reference sequence
  employed by GISAID (EPI\_ISL\_402124). Learn more at
  \url{https://gisaid.org/WIV04}.
\end{itemize}

\newpage

\subsection{BEAST modeling details for real data investigation:
Washington
State}\label{beast-modeling-details-for-real-data-investigation-washington-state}

\begin{table}

\caption{\label{tbl-beast-details}BEAST modeling details for each real
data scenario for Washington State samples reported to GISAID. All
unnoted specifications were left as BEAST defaults.}

\centering{

\centering
\begin{tabular}[t]{>{\raggedright\arraybackslash}p{3.5cm}>{\raggedright\arraybackslash}p{3.5cm}>{\raggedright\arraybackslash}p{3.5cm}>{\raggedright\arraybackslash}p{3.5cm}}
\toprule
  & Full Scenario & Observed Scenario & Truncated Scenario\\
\midrule
Data & Sampled before 2021-08-01, inclusive & Reported before 2021-08-01, inclusive & Sampled before 2021-06-21, inclusive\\
\addlinespace[0.3em]
\multicolumn{4}{l}{Models}\\
\hspace{1em}Substitution & HKY & HKY & HKY\\
\hspace{1em}Clock type & strict & strict & strict\\
\hspace{1em}Coalescent & Bayesian Skygrid & Bayesian Skygrid & Bayesian Skygrid\\
\hspace{1em}\hspace{1em}\# of parameters & 50 & 50 & 50\\
\hspace{1em}\hspace{1em}Last transition & 1.63 years & 1.52 years & 1.6 years\\
\addlinespace[0.3em]
\multicolumn{4}{l}{Priors}\\
\hspace{1em}Kappa & LogNormal(1, 1.25) & LogNormal(1, 1.25) & LogNormal(1, 1.25)\\
\hspace{1em}Frequencies & Dirichlet(1, 1) & Dirichlet(1, 1) & Dirichlet(1, 1)\\
\hspace{1em}Clock rate & Unif(3e-4, 1.1e-3) & Unif(3e-4, 1.1e-3) & Unif(3e-4, 1.1e-3)\\
\hspace{1em}Root height & None (tree prior only) & None (tree prior only) & None (tree prior only)\\
\hspace{1em}Skygrid precision & Gamma(0.001, 1000) & Gamma(0.001, 1000) & Gamma(0.001, 1000)\\
\addlinespace[0.3em]
\multicolumn{4}{l}{MCMC options}\\
\hspace{1em}Chain length & 2e+07 & 2e+07 & 2e+07\\
\hspace{1em}Burn in & 2500000 & 2500000 & 2500000\\
\hspace{1em}Log every & 2000 & 2000 & 2000\\
\hspace{1em}Seed & - & - & -\\
\bottomrule
\end{tabular}

}

\end{table}%

\begin{figure}[H]

{\centering \includegraphics[width=\textwidth]{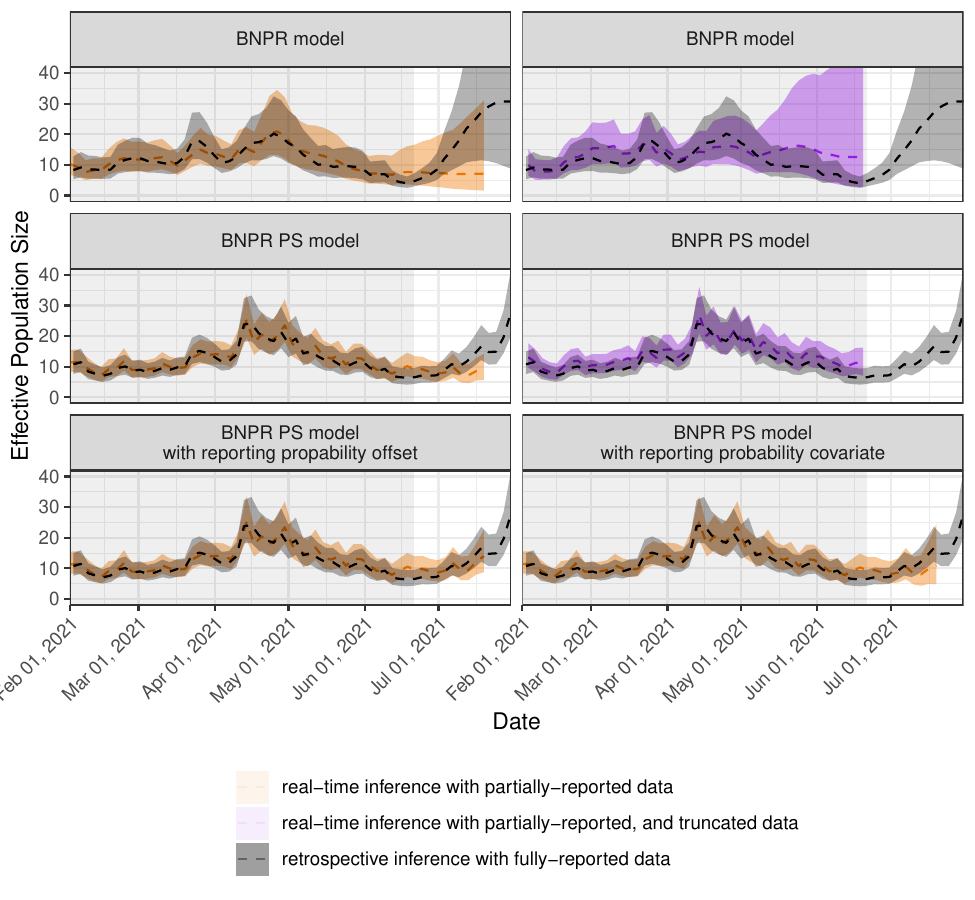}

}

\caption{Comparison of phylodynamic estimation methods of effective
population size trajectory for Washington State SARS-CoV-2 sequences.}

\end{figure}%

\end{document}